\documentclass[%
 reprint,
 amsmath,amssymb,
 aps,
prl,
]{revtex4-2}

\usepackage{graphicx}
\usepackage{dcolumn}
\usepackage{bm}
\usepackage[caption=false]{subfig}
\usepackage{hyperref}
\hypersetup{unicode=true, colorlinks=true, linkcolor=blue, citecolor=blue}

\begin{document}
\setcounter{secnumdepth}{3}
\renewcommand{\thesection}{\Roman{section}}

\preprint{APS/123-QED}
\title{Stochastic Models of Neuronal Growth}

\author{Cristian Staii}%
\email{cstaii01@tufts.edu}
\affiliation{Department of Physics and Astronomy, Tufts University, Medford, MA, 02155, USA}

\date{\today}

\begin{abstract}

Neuronal circuits arise as axons and dendrites extend, navigate, and connect to target cells. Axonal growth, in particular, integrates deterministic guidance from substrate mechanics and geometry with stochastic fluctuations generated by signaling, molecular detection, cytoskeletal assembly, and growth-cone dynamics. A comprehensive quantitative description of this process remains incomplete.
We review stochastic models in which Langevin dynamics and the associated Fokker--Planck equation capture axonal motion and turning under combined biases and noise. Paired with experiments, these models yield key parameters, including effective diffusion (motility) coefficients, speed and angle distributions, mean-square displacement, and mechanical measures of cell--substrate coupling, thereby linking single-cell biophysics and intercellular interactions to collective growth statistics and network formation. We further couple the Fokker--Planck description to a mechanochemical actin--myosin--clutch model and perform a linear stability analysis of the resulting dynamics. Routh--Hurwitz criteria identify regimes of steady extension, damped oscillations, and Hopf bifurcations that generate sustained limit cycles.
Together, these results clarify the mechanisms that govern axonal guidance and connectivity and inform the design of engineered substrates and neuroprosthetic scaffolds aimed at enhancing nerve repair and regeneration.

\end{abstract}

\maketitle

\section{Introduction}

The human brain comprises an immense network of neurons whose axons and dendrites (collectively known as neurites) establish long range, highly specific connections during development \cite{Striedter2016, Arimura2007, Huber2003, Lowery2009, Azevedo2009}. Axons often extend over distances of tens to hundreds of cell diameters to reach appropriate targets, a process directed by the growth cone, a dynamic sensor and actuator complex at the axon tip that integrates biochemical, mechanical, and topographical cues to drive directed motion \cite{Franze2010, Kiryushko2004, Lowery2009, Goodhill2003, Franze2013}. Guidance signals include diffusible molecules (e.g., Netrins, Slits, Semaphorins) and substrate bound factors (Ephrins, extracellular matrix components, adhesion molecules), together with physical inputs such as stiffness, geometry, and electric fields \cite{Franze2013, Huber2003, Moeendarbary2014, Riveline2001, Fivaz2007, Samuels1996, Toriyama2010, Franze2020, Alert2020}. Notably, growth cones navigate heterogeneous microenvironments with high precision, continuously probing and updating their trajectory \cite{Franze2009, Franze2020, Lowery2009, Takano2019}. These processes ultimately wire circuits that enable reflexes, learning, attention, and memory.

A central challenge is to translate this complexity into predictive, quantitative laws. Axonal extension is inherently noisy: ligand detection near the single-molecule limit, stochastic reaction networks, intermittent adhesion engagement, and fluctuating cytoskeletal remodeling all introduce variability at the scale of the growth cone. At the same time, cells exploit regulatory feedback to stabilize motion and amplify relevant cues \cite{El-Samad2021, Collier1996, Takano2019, Oliveri2022, Descoteaux2022}. In the molecular clutch model, actin polymerization at the leading edge, retrograde flow driven by myosin-II, and dynamic coupling to substrate adhesions (integrins, cadherins) together determine traction forces and growth cone advance \cite{Lowery2009, Franze2010, Huber2003, Medeiros2006, Lilja2018, Polackwich2015, Pouwels2012, Jurchenko2015, Buskermolen2020, Hyland2014, Koch2012, Kumarasinghe2022}. Positive feedback reinforces forward motion and alignment, whereas negative feedback damps fluctuations and prevents uncontrolled responses \cite{Lowery2009, Takano2019, Fivaz2007, Riveline2001}. These considerations strongly motivate modeling frameworks based on \emph{stochastic processes}, such as Langevin and Fokker–Planck equations, that explicitly couple deterministic axonal guidance (drift) with random fluctuations (diffusion), and inherently incorporate feedback-modulated noise. In these models, clutch‑mediated force transmission contributes to the drift term, while polymerization of the cytoskeleton, adhesion turnover, and signaling noise set the diffusion term.

Stochastic differential equations (SDE) provide a compact way to encode this interplay: drift fields represent guidance and feedback-regulated tendencies (e.g., alignment torques induced by patterned substrates), while diffusion coefficients capture intrinsic and extrinsic noise whose magnitude may depend on the local microenvironment. From a modeling perspective, axonal trajectories reflect the interplay of (i) deterministic biases imparted by chemical gradients, substrate mechanics, and geometry, and (ii) stochastic fluctuations arising from receptor binding, signaling cascades, adhesion engagement, and cytoskeletal remodeling.  This formalism supports parameter inference from data and enables hypothesis tests about mechanism (e.g., clutch-mediated traction vs. gradient sensing) using likelihood-based or information-theoretic criteria \cite{El-Samad2021, Collier1996, pearson2011modeling, DeGennes2007, Murray1993, Simpson2009, Oliveri2022, Oliveri2021, Jakobs2020}.

\begin{figure*}[ht]
    \centering
    \includegraphics[width=13cm]{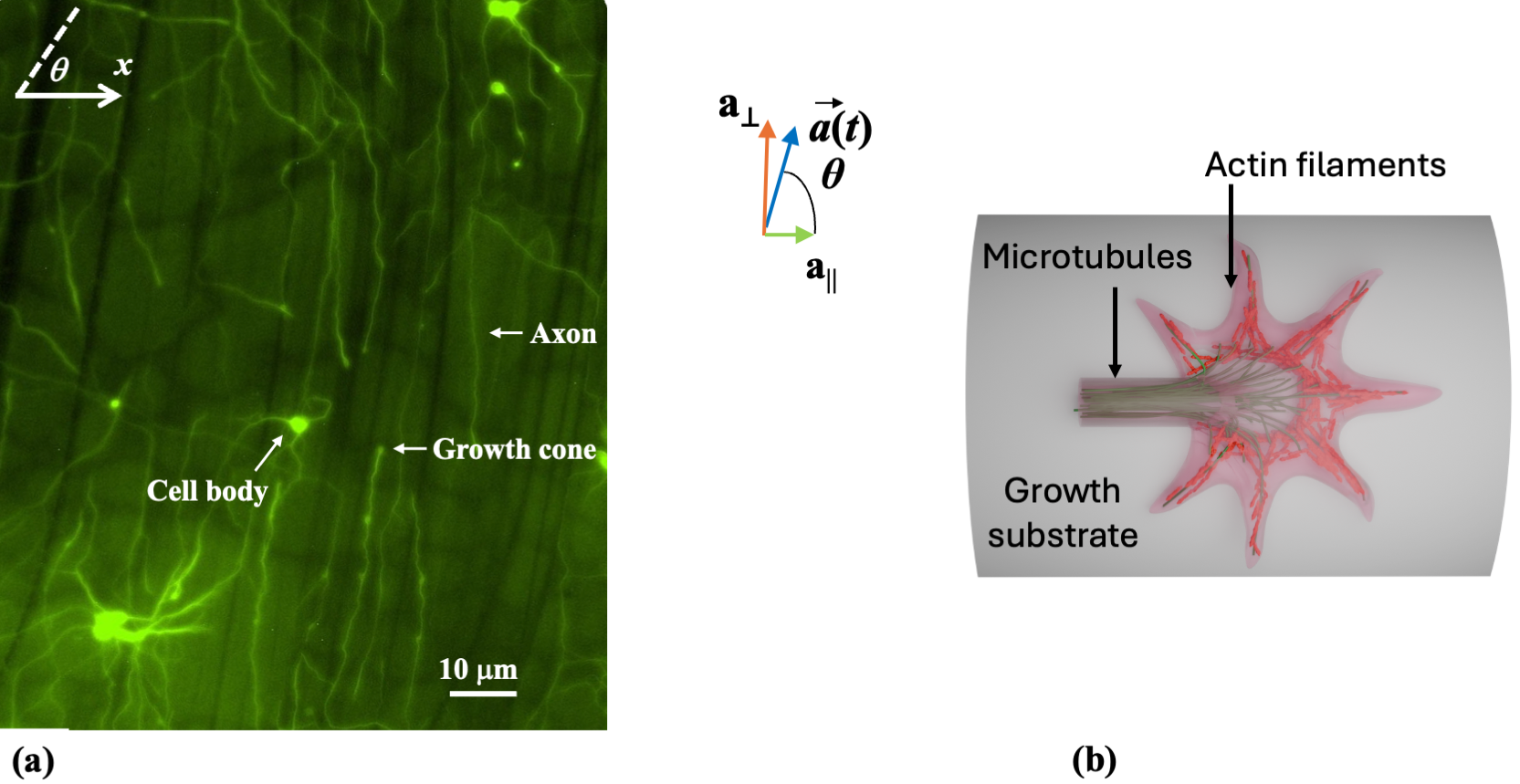}
    \caption{:
(\textbf{a}) Fluorescence image showing examples of axonal growth for cortical neurons cultured on a poly-D-lysine (PDL) coated surface with periodic micropatterns. Cortical neurons typically grow a long process (axon) and several minor processes (dendrites). Axonal growth is directed by the growth cone. The axons is identified by its morphology and the growth cone is identified as the tip of the axon. The angular coordinate $\theta$ used in this paper is also defined in this figure. The figure inset shows the parallel and perpendicular components of the acceleration in the reference frame used in the paper (see main text). (\textbf{b}) Schematic representation of the growth cone structure, illustrating the major cytoskeletal components: actin filaments and microtubules. The actin filaments are mechanically linked to the growth substrate through point contacts formed by transmembrane cell adhesion molecules, such as integrins and cadherins. The interaction among integrins, adhesion proteins, actin filaments  and microtubules generates traction forces that promote the advancement  of the growth cone.   \label{fig1}}
\end{figure*}

Recent advances in microfabrication and microfluidics provide controlled \emph{in vitro} platforms to calibrate and test theoretical models. Engineered culture systems allow independent tuning of biochemical, mechanical, and geometric cues and reveal strong stiffness dependence of axonal elongation and robust guidance by patterned topographies \cite{Teixeira2020, kundu2013superimposed, gladkov2017design, Hyland2014, Jurchenko2015, ishihara2011assay, Descoteaux2022, Spedden2012Elasticity, Koch2012, CastroDominguez2023, Pfister2011}. For example, on grooved or anisotropic substrates, axons display biased, persistent motion and enhanced alignment (Figure~\ref{fig1} shows an example of alignment for axonal growth on micropatterned surfaces). These are signatures naturally captured by drift–diffusion, velocity-jump, or biased, persistent random walk models \cite{Rizzo2013, Basso2019, Yurchenko2021, Staii2023, Yurchenko2019, Vensi2019, Descoteaux2022, Oliveri2021, Athamneh2017}. Such descriptions yield directly measurable predictions for speed and turning angle distributions, mean-square displacements, velocity and angular correlation functions, and motility coefficients, thereby linking single-cell biophysics (cytoskeletal dynamics, adhesion kinetics) to ensemble-level trajectory statistics and, ultimately, circuit-level connectivity.

In our previous studies, we have demonstrated that cortical neurons grown on poly-D-lysine (PDL) coated substrates with periodic micropatterns exhibit strong alignment with the surface features \cite{Rizzo2013, Basso2019, Yurchenko2021, Yurchenko2019, Staii2023, Staii2024}. This behavior is consistent with an effective, substrate-induced deterministic torque and with feedback that modulates adhesion and cytoskeletal dynamics \cite{Descoteaux2022, Kumarasinghe2022}. We quantified axonal speeds, angular distributions, autocorrelation functions, diffusion (cell motility) coefficients, and cell–surface interaction forces. We have also extracted mechanical parameters such as elastic and bending modulus relevant to shape control during guidance \cite{Descoteaux2022, Yurchenko2019, Spedden2012Elasticity}. Datasets obtained from these experiments are well suited for building stochastic models, testing scaling predictions (e.g., mean squared displacement vs. time), and quantifying the dependence of drift and diffusion on controllable environmental cues.

A quantitative, stochastic account of axonal guidance is also essential for engineering growth-permissive microenvironments and for therapeutic strategies in nerve repair and neurodegeneration. Insight into how feedback and noise shape axonal outgrowth can inform the design of neuroprosthetic scaffolds and bioinspired platforms that promote targeted regeneration and functional reconnection \cite{Pfister2011, Holland2015, Bayly2014, DeRooij2018a, DeRooij2018b, Mahar2018, Teixeira2020, Ahmadzadeh2015, Montanino2018}. Because stochastic frameworks return experimentally measurable parameters (drift strengths, diffusion coefficients, correlation times), they provide actionable targets for materials design (e.g., stiffness, biochemical composition, pattern periodicity) and for pharmacological modulation of adhesion and cytoskeletal dynamics. For example, increasing pattern anisotropy strengthens orientation drift, whereas stabilizing adhesions reduces effective diffusion by lengthening correlation times \cite{Vensi2019, Yurchenko2019, Descoteaux2022}.

This paper advances a stochastic perspective on neuronal growth. We present evidence that mechanical and biochemical guidance cues operate through feedback-regulated mechanisms to produce biased yet noisy trajectories, and we formalize this behavior with Langevin and Fokker–Planck models. We show how to use these models to interpret experimental data and to extract diffusion coefficients, drift fields, turning-angle and speed distributions, and parameters quantifying cell–substrate coupling. Together, these results demonstrate that stochastic models are not merely descriptive: they provide a compact, predictive framework linking intracellular dynamics to axonal guidance and network formation in complex, fluctuating environments. We begin with a general overview of mathematical models of neuronal growth. We then formulate Langevin and Fokker–Planck models with nonlinear drift terms, analyze growth angle and velocity distributions, and calibrate parameters using trajectory data on micropatterned substrates. Finally, we introduce a stochastic mechanochemical model that couples a Fokker–Planck description of growth cone position to explicit dynamics of actin polymerization, myosin-II–driven contraction, and point-contact adhesions. We show that regulated feedback loops between polymerization, contractility, and adhesion govern growth velocity and adhesion-dependent traction, collectively producing steady extension, damped oscillations, and limit cycles. A linear stability analysis of the coarsed-grained dynamics delineates the parameter regimes associated with each behavior and clarifies how these coupled feedbacks jointly tune axonal outgrowth.

\section{Mathematical Modeling of Axonal Growth }

Modeling frameworks for axonal growth range from phenomenological descriptions of trajectories to mechanistic models of growth cone biophysics. Early work treated growth-cone motion as a random walk, asking whether observed elongation–retraction dynamics could be explained without invoking detailed intracellular mechanisms. For example, Katz and colleagues showed that net advance and retraction are, to first approximation, consistent with an uncorrelated random walk \cite{Katz1984}, whereas Odde and collaborators reported short-time correlations between extension and subsequent retraction on minute timescales \cite{Odde1996}. In parallel, Buettner and co-workers extracted probabilistic rules for filopodial extension–retraction from time-lapse imaging and formalized these rules into a stochastic model \cite{Buettner1994, Buettner1995}. At the level of chemical sensing, the Goodhill group developed statistical models of cue–receptor binding at the growth cone \cite{Simpson2009}, deriving constraints set by gradient shape and noise on detectability and steering \cite{Goodhill1998}, and showing that spatial sensing outperforms purely temporal strategies across experimentally relevant concentration ranges \cite{Goodhill1999}. Katz and Lasek further identified constraints required to obtain ordered axonal ensembles from simple random-walk processes \cite{Katz1985}. These studies established the utility of stochastic kinematic descriptions and sensing-theory bounds for interpreting growth trajectories.

Because biochemistry and mechanics are multiscale, mechanistic modeling has concentrated on tractable subsystems or controlled environments. Segev and Ben-Jacob modeled self-wiring in diffusing guidance fields and used graph-theoretic metrics to compare emergent networks with experiments \cite{Segev2000}. Van Ooyen’s group simulated multiple axons navigating domains with overlapping guidance cues \cite{Krottje2007}. At the subcellular level, Mogilner and Rubenstein developed a mechanical theory of filopodial architecture to infer optimal length and stability \cite{Mogilner2005}. Padmanabhan and Goodhill incorporated a molecular feedback loop in cytoskeletal control pathways that yields unimodal or bistable outgrowth depending on point contact adhesion assembly rates. Combined with a stochastic angular process, this produces a random walk with rest model in which advance and pausing reflect the state of the internal switch \cite{Padmanabhan2018}. Reduced compartmental models have also been proposed to predict growth cone responses to externally imposed gradients \cite{Lin2020}. Collectively, these contributions link intracellular regulation (adhesion, cytoskeletal turnover, signaling feedback) to mesoscopic motion under explicit biophysical assumptions.

A complementary line of work casts axonal guidance as \emph{stochastic transport} governed by SDEs, with deterministic drift encoding biases from chemical gradients, substrate mechanics, or geometry, and diffusion capturing intrinsic and extrinsic fluctuations (receptor noise, reaction networks, adhesion engagement, and cytoskeletal remodeling). Simulating these SDEs yields probability densities over position and orientation, enabling direct comparison with ensemble statistics (speed and turning-angle distributions, mean-square displacement, correlation functions) and testable predictions for competing biophysical mechanisms. Using such approaches, Hentschel and van Ooyen reproduced axonal bundling, guidance, and subsequent debundling in combined attractant–repellent fields \cite{Hentschel1999}. Maskery and Shinbrot used simulations based on SDE to estimate minimum detectable gradients under realistic noise \cite{Maskery2005}. Pearson and colleagues obtaining baseline trajectory geometries for growth in cue-free environments,  \cite{pearson2011modeling}. Goodhill and collaborators coupled ligand binding with filopodial dynamics to generate guided trajectories in imposed gradients \cite{Goodhill2004}. At the subcellular scale, Betz and co-workers used SDE analysis to lamellipodial fluctuations, showing that observed bimodality emerges from actin-driven bistability \cite{Betz2006}. These studies illustrate how drift–diffusion models provide compact, data-driven links between microenvironmental statistics and growth cone kinematics.

Beyond single-axon descriptions, agent-based and network-level models incorporate interaction rules (for example, fasciculation/defasciculation, competition for cues) and domain topology. With local sensing and adhesion rules, simulations recover collective alignment, bundle formation, and target selection in patterned or heterogeneous landscapes \cite{Segev2000, Krottje2007}. In such settings, stochasticity is not merely noise but a resource: fluctuations enable escape from local traps, exploration of alternative routes, and sensitivity to weak biases, while feedback modulates noise levels to stabilize chosen paths \cite{El-Samad2021, Oliveri2022, Cheng2025}.

\textit{Novelty and Scope of Contribution}. Existing models of axonal growth have typically focused either on kinematic trajectory statistics or on detailed mechanochemical regulation, but rarely on their explicit integration. Gradient-sensing and filopodial models in the tradition of Goodhill and co-workers quantify how noisy cue--receptor binding constrains guidance and steering, while multiscale mechanical descriptions in the spirit of Franze and collaborators emphasize substrate stiffness, tension, and curvature as control parameters for neurite outgrowth. Likewise, van Ooyen and co-workers developed stochastic frameworks for axonal guidance and bundling in prescribed cue fields, and clutch-based mechanochemical models have clarified how actin polymerization, myosin contractility, and adhesion dynamics jointly regulate traction at the growth cone. Building on these advances, the present work combines a Langevin/Fokker--Planck description of growth cone motion with an explicit, coarse grained actin--myosin--clutch model, thereby linking experimentally accessible drift and diffusion coefficients directly to intracellular feedback variables.

\section{Langevin and Fokker-Planck Formalisms for Modeling Axonal Dynamics}

Axonal growth arises from the interplay between deterministic and stochastic components of growth cone motility. Deterministic biases emerge, for example, from preferred orientations imposed by substrate geometry, whereas the stochastic contributions originate from cytoskeletal polymerization (actin and microtubules), intracellular signaling, detection of low-concentration cues, biochemical reactions, and the formation and turnover of lamellipodia and filopodia \cite{Striedter2016, Arimura2007, Huber2003, Lowery2009, Azevedo2009, Franze2010, Kiryushko2004, Goodhill2003}. Because of this interplay, single-neuron trajectories are not deterministically predictable. However, ensemble behavior can be captured by probability densities governed by the associated SDEs. In particular, Langevin dynamics and the associated Fokker-Planck equation (FPE) provide a compact framework for modeling axonal dynamics as drift–diffusion processes that integrate guidance cues with noise \cite{Hentschel1999, Goodhill2004, Maskery2005, pearson2011modeling, Betz2006}. 

These stochastic models are particularly powerful when calibrated and validated against controlled \emph{in vitro} measurements. By fitting SDE/FPE parameters to axonal trajectories on engineered substrates, one can extract effective drift fields (mechanical guidance strengths, alignment torques), diffusion (cell motility) coefficients, and correlation times, then test scaling laws such as mean-squared displacement (MSD) growth or velocity and angular correlations. For example, in our prior work, we have shown that cortical neurons grown on PDL-coated glass exhibited dynamics consistent with an effective V-shaped potential that regulates growth rates \cite{Rizzo2013}. On ratchet-like, tilted-nanorod (nano-ppx) surfaces, axons aligned along a preferred direction due to a substrate-induced deterministic torque. We have measured angular distributions and drift–diffusion coefficients that quantify this bias \cite{Spedden2014Asymmetry, beighley2012alignment}. In a separate set of experiments, we showed that periodic geometrical patterns impart strong directional bias to axonal growth (Figure~\ref{fig1}) \cite{Basso2019, Yurchenko2021, Yurchenko2019, Staii2023, Staii2024, Kumarasinghe2022, Descoteaux2022}. We have measured growth cone speeds, velocity autocorrelations, axonal orientation distributions, diffusion coefficients, and neuron-substrate traction forces. These examples show how stochastic transport models serve as a unifying language to compare disparate conditions (chemical gradients, mechanical stifness, geometrical anisotropy) and to map microenvironmental control parameters to observable path statistics.

\begin{figure*}[ht]
 \centering
    \includegraphics[width=13 cm]{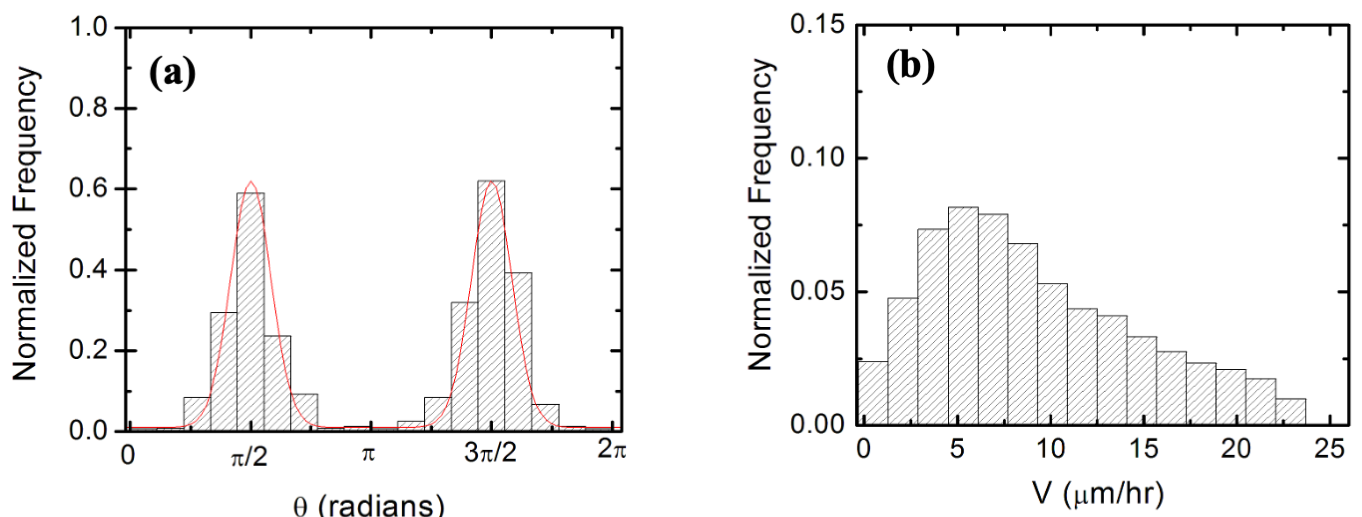}
    \caption{(\textbf{a}) Example of normalized experimental angular distributions for axonal growth for neurons cultured on micropatterned surfaces. The vertical axis (labeled Normalized Frequency) represents the ratio between the number of axonal segments growing in a given direction and the total number N of axon segments. Each axonal segment is of 10 $\mu m$ in length. The plot shows the angular distribution for  $N$ = 510 different axon segments (total of 67 axons). The data shows that the axons display strong directional alignment along the surface patterns (peaks at $\theta = \pi/2$  and $\theta = 3\pi/2$. The continuous red curve represents a fit to the data with the Fokker-Planck model discussed in the text (\textbf{b}) Normalized speed distributions measured for the growth cones of the axons shown in (a). Data points were collected at $t = 32$ hr after neuron plating.  \label{fig2}}
\end{figure*} 

\textit{3.1 Langevin equation for axonal growth.}
In previous work \cite{Vensi2019} we have shown that axonal dynamics on uniform glass surfaces is described by an Ornstein-Uhlenbeck (Brownian) process, defined by a linear Langevin equation for the velocity $\Vec{V}$:
\begin{equation}
\frac{d\vec{V}}{dt} = -\,\gamma_g\,\vec{V} \;+\; \vec{\Gamma}(t)
\label{eq:A1}
\end{equation}
with constant damping $\gamma_g$ and Gaussian white noise $\vec\Gamma(t)$.

From Equation~\eqref{eq:A1} we calculate the mean square axonal length and the velocity autocorrelation function \cite{Vensi2019}.  By comparing the theoretical predictions with the experimentally measured distributions for these parameters we can extract the two fundamental parameters that characterize the motion of axons on glass surfaces: the diffusion coefficient $D$ and the characteristic time for the exponential decay of the velocity autocorrelation function $\tau_g=1/\gamma_g$. For cortical neurons grown on PDL coated glass these parameters are : $D = (16 \pm 2)\,\mu\mathrm m^{2}\,\mathrm h^{-1}$ and $\gamma_g = (0.1 \pm 0.05)\,\mathrm h^{-1}$ \cite{Vensi2019}. 

Axonal dynamics on micropatterned surfaces is described by a non-linear Langevin equation:
\begin{equation}
\frac{d\vec {V}}{dt} \;\equiv\; \vec a(\vec V,t)
\;=\; \vec a_{\ d}(\vec V,t) \;+\; \vec\Gamma(t),
\label{eq:A2}
\end{equation}
where $\vec a_d$ is the deterministic component and $\vec\Gamma(t)$ the stochastic term. The acceleration of axons is decomposed into a component parallel to the direction of motion  $\vec a_{d, \parallel}(\vec V,t)$, and a component perpendicular to this direction $\vec a_{d, \perp}(\vec V,t)$ (inset in Figure~\ref{fig1}). A separate analysis of the two motions leads to the following non-linear Langevin equations for the two components of the acceleration \cite{Vensi2019}:
\begin{equation}
a_{\parallel}(V,t) \equiv \frac{d \vec V_{\parallel}}{dt}
= a_0 |\mathrm{sin}(\theta)| \;-\; \gamma_1\,V \;-\; \gamma_2\,V^2 \;+\; \Gamma_{\parallel}(t)
\label{eq:A3}
\end{equation}
\begin{equation}
a_{\perp}(V,t) \equiv \frac{d \vec V_{\perp}}{dt}
= a_1 \,\cos\theta \;+\; \Gamma_{\perp}(t)
\label{eq:A4}
\end{equation}
Here, $\theta$ is growth angle, $V$ is the growth cone speed, and $a_0,a_1,\gamma_1,\gamma_2$ are velocity-independent parameters that characterize axonal dynamics on substrates with periodic geometries. $\Gamma_{\parallel},\Gamma_{\perp}$ are independent Gaussian white noises for parallel and perpendicular growth. We have shown that all these parameters are experimentally measurable \cite{Vensi2019}.

Equations~\eqref{eq:A3}-\eqref{eq:A4} show that the axonal dynamics on surfaces with periodic geometries is described by non-linear Langevin equations, involving quadratic velocity terms and non-zero coefficients for the angular orientation of the growing axon. There are some very important consequences for axonal growth that follow from this type of dynamics. In particular, Equations~\eqref{eq:A3}-\eqref{eq:A4} show angular alignment of axonal growth on micropatterned surfaces. The magnitude of the perpendicular acceleration $a_{\perp}$ has a maximum value when the direction of axonal growth is perpendicular to the surface pattern (i.e., for $\theta=\pi/2, \ 3\pi/2$ in Figures ~\ref{fig1}, and ~\ref{fig2}), and it equals zero when the axon grows along the pattern ($\theta=0, \ \pi$). This shows that the perpendicular component of acceleration $a_{\perp}$ tends to align the growth cone along the direction of the pattern. The net effect is that of a deterministic torque (quantified by the parameter $a_{1}$) which rotates the growth cone towards the surface geometrical pattern. Figure~\ref{fig2} shows examples of angular (Figure~\ref{fig2}a) and speed (Figure~\ref{fig2}b) distributions for axonal growth on a surface with periodic micropatterns.

Another prediction of the model described by Equations~\eqref{eq:A3}-\eqref{eq:A4} is that the growth cone reaches a terminal speed along the direction of the pattern, which can be found from the condition that the average acceleration in Equation~\eqref{eq:A3} equals zero. This gives the following analytic expression for the terminal speed of the growth cone:
\begin{equation}
V_{\rm ter} \;=\; \sqrt{\frac{a_0}{\gamma_2}\cdot|{\sin\theta}|+\frac{\gamma_1 ^2}{4\gamma_2 ^2}} -\frac{\gamma_1}{2\gamma_2}
\label{eq:A5}
\end{equation}

Equation~\eqref{eq:A5} has a number of features that can be tested experimentally.  First, the growth cones reaches terminal speed only for growth angles $\theta \neq 0$. In addition, the terminal speed depends only on the ratios of the growth parameters $a_o/\gamma_2$, and $\gamma_1/\gamma_2$ which ultimately depend on the substrate geometry. 
Another important consequence of the non-linear Langevin Equations~\eqref{eq:A3}-\eqref{eq:A4} is that axonal growth displays a cross-over from Brownian motion at earlier to a supper-diffusion regime at later times. The supper-diffusive dynamics is characterized by non-Gaussian speed distributions and power law increase of the axonal mean square length with time \cite{Yurchenko2019, Staii2024}. From a biological perspective, the observed transition between the diffusive to super-diffusive axonal motion suggests long-range spatial and temporal correlations in the underlying dynamics \cite{Yurchenko2019}. 

\textit{3.2 Fokker-Planck equations for axonal growth.}
In a series of papers \cite{Vensi2019, Yurchenko2019, Basso2019, Descoteaux2022} we have shown that the axonal dynamics on surfaces with periodic geometries is completely described by the following system of Fokker-Planck  equations.
\paragraph{(a) Fokker--Planck equation for spatial probability $P(\vec{r,t)}$ (Smoluchowski form):}
\begin{equation}
\partial_t P(\vec r,t)
= D\,\nabla^2 P(\vec r,t) \;+\; \frac{1}{\gamma}\,\nabla\!\cdot\!\bigl(P(\vec r,t)\,\nabla U(\vec r)\bigr)
\label{eq:A6}
\end{equation}
with diffusion (motility) coefficient $D$, damping $\gamma$, and effective potential $U(\vec r)$. The 1D stationary solution along micropatterns is \cite{Descoteaux2022}:
\begin{equation}
\begin{aligned}
P(\vec{r},t) &= A\,p(x,t),\qquad
p(x)=\frac{1}{Z}\exp\!\left[-\frac{\gamma}{D}\,U_{\rm eff}(x)\right],\\
U_{\rm eff}(x) &= U_{\rm sub}(x)+U_{\rm fb}(x)+U_{\rm int}(x).
\end{aligned}
\label{eq:A7}
\end{equation}
where $A,Z$ are normalization constants, $U_{\rm sub}$ is the external potential imposed by the substrate geometry, $U_{\rm fb}$ the feedback potential, and $U_{\rm int}$ accounts for neuron--neuron interactions. The form of these three potentials has been studied in reference \cite{Descoteaux2022}.

\paragraph{(b) Fokker--Planck equation for the speed distribution $P(v,t)$:}
\begin{equation}
\partial_t P(v,t)
= \partial_v\!\Big[\gamma_v\,(v-\bar v)\,P(v,t)\Big]
\;+\; \frac{\sigma^2}{2}\,\partial_v^2 P(v,t)
\label{eq:A8}
\end{equation}
The solution of this equation for the stationary speed distribution is:
\begin{equation}
P(v)=B\,\exp\!\left[-\,\frac{\gamma_v}{\sigma^2}\,\bigl(v-\bar v\bigr)^2\right],
\qquad \int_{0}^{\infty}P(v)\,dv=1
\label{eq:A9}
\end{equation}
with damping $\gamma_v=1/\tau$ (relaxation time $\tau$), mean speed $\bar v$, Gaussian noise strength $\sigma$, and normalization constant $B$.

\paragraph{(c) Fokker--Planck equation for the angular probability $P(\theta, t)$:}
\begin{equation}
\partial_t P(\theta,t)
= \partial_\theta\!\Big[-\,\gamma_\theta\cdot\cos\theta\cdot P(\theta,t)\Big]
\;+\; D_\theta\,\partial_\theta^2 P(\theta,t)
\label{eq:A10}
\end{equation}
The solution of this equation for the stationary angular distribution is:
\begin{equation}
P(\theta) = C\,\exp\!\Big[\frac{\gamma_{\theta}}{D_{\theta}}|\sin\theta|\Big],
\quad \int_{0}^{2\pi}P(\theta)\,d\theta=1
\label{eq:A11}
\end{equation}
In Equations~\eqref{eq:A10}-\eqref{eq:A11} $P(\theta,t)$ is the probability distribution for the growth angle $\theta$,  $C$ is a normalization constant,  $D_{\theta}$ represents the effective angular diffusion coefficient, and $\gamma_{\theta}\cos \theta(t)$ corresponds to a “deterministic torque” representing the tendency of the growth cone to align with the preferred growth direction imposed by the surface geometry. The absolute value $\lvert \sin\theta \rvert$ in Equation~\eqref{eq:A11} reflects the symmetry of axonal growth around the $x$ axis: the angular distributions centered at $\theta$ are symmetric with respect to the directions $\theta$ and $\pi-\theta$ (Figure~\ref{fig2}). This is a consequence of the fact that there is no preferred direction along the substrate micropattern. We also note that the deterministic torque has a maximum value if the growth cone moves perpendicular to the surface patterns ($\theta=0$ or $\theta=\pi$), in which case the cell--surface interaction tends to align the axon with the surface pattern. The torque is zero for an axon moving along the micropattern.

In previous work, we have used the model given by Equations~\eqref{eq:A6}-\eqref{eq:A11} to extract key dynamical parameters of axonal motion. Typical values obtained for the growth parameters are: diffusion coefficient $D=(22\pm4)\,\mu\mathrm{m}^2/\mathrm{hr}$, coefficient for the ``deterministic'' alignment torque $\gamma_{\theta}=(0.13\pm0.04)\,\mathrm{hr}^{-1}$, and characteristic time for axonal alignment $\tau=(5.1\pm0.8)\,\mathrm{hr}$. We have performed a detailed analysis of how these parameters depend on the type of substrate, growth time, and chemical modification of the neurons \cite{Basso2019, Yurchenko2021, Descoteaux2022}. These results show that the dynamics of the ensemble of axons can be described phenomenologically if each growth cone is modeled as an automatic controller with a closed feedback loop \cite{Descoteaux2022}. Growth alignment is fully determined by the surface geometry, and the distance between  micropatterns plays the role of a control parameter. In particular, we have performed experiments which demonstrate that the disruption of cytoskeletal dynamics through neuronal treatment with different chemical compounds alters the feedback loop of the cellular controller \cite{Basso2019, Yurchenko2021, Descoteaux2022}.

\section{Mechanical Beam Model of Axons}
The phenomenological models discussed in the previous sections form a basis for quantifying cell-cell and cell-surface interactions, and ultimately for describing how the formation of neuronal network emerges from collective biophysical processes of single cells. In particular, the Fokker-Planck dynamics can be justified by a simple mechanical model that takes into account the cell-substrate interactions \cite{Descoteaux2022}. The model considers the bending-induced strained sustained by the axon while growing on the semi-cylindrical pattern of radius $R$: axonal adhesion to the surface leads to axonal bending, which in turn leads to increased mechanical strain energy in the axon cytoskeleton. The mechanical strain energy $E$ depends on the axon bending modulus $F$, and the local surface curvature $\kappa(\theta,R)$ \cite{Descoteaux2022} :
\begin{equation}
E \;=\; \frac{F}{2}\,\kappa^2(\theta,R)
\label{eq:A12}
\end{equation}
In the case of axonal growth on the micropatterned surfaces, the curvature of an axon  around the cylindrical pattern of radius $R$ is given by: 
\begin{equation}
\kappa(\theta,R) \;=\; \frac{|\cos\theta|}{R}
\label{eq:A13}
\end{equation}
For the stationary growth described by the Fokker-Planck model one can assume a Boltzmann-type distribution for the probability of axon growing in a given direction: 
\begin{equation}
P(\theta)\;=\; A_1\,\exp\!\left(-\frac{E}{E_0}\right) \ = A_1\,\exp\!\left(-\frac{F}{E_0\cdot R^2}\cdot\cos^2\theta \right) \ 
\label{eq:A14}
\end{equation}
where $E_0$ is the characteristic energy scale for axonal bending, and $A_1$ is an overall normalization constant. 
By comparing the solution of this simple mechanical beam model (Equation~\eqref{eq:A14}) with the stationary solutions of the Fokker--Planck model (Equations~\eqref{eq:A9} and \eqref{eq:A11}), and using the experimentally measured values for the radius \(R\) of curvature of the micropattern and the growth parameters \(D_{\theta}\) and \(\gamma_{\theta}\), one can extract the bending modulus of the axon. Typical values for the bending modulus are \(F \approx 23~\mathrm{J}\,\mu\mathrm{m}^{2}\) for untreated neurons, and \(F \approx 17~\mathrm{J}\,\mu\mathrm{m}^{2}\) for neurons in which the cytoskeletal dynamics was inhibited by chemical modification \cite{Descoteaux2022}.

These results indicate that axonal stiffness and substrate curvature can act jointly to direct axonal growth. The framework can be extended to include explicit dependencies of the growth parameters on biomechanical and geometric guidance cues, such as substrate geometry and stiffness, as well as on externally applied forces. For example, a proposed model for the cooperative motion of close‑packed cell ensembles treats contractile forces and effective cellular polarization as internal variables that generate waves of collective migration \cite{Notbohm2016}. Continuum mechanical models that couple cell–substrate interactions to cellular biomechanical properties have likewise been proposed \cite{Banerjee2011, Dokukina2010, Barnhart2011, Rubinstein2009}. The parameters in these models are accessible experimentally via combined Atomic Force Microscopy (AFM) and Traction Force Microscopy (TFM) measurements \cite{Spedden2012Elasticity, Kumarasinghe2022, Descoteaux2022}.

\begin{table*}[ht]
    \centering
    \small
    \begin{tabular}{@{}c@{\hspace{24pt}}c@{\hspace{24pt}}c@{\hspace{24pt}}c@{}}
        \hline\hline
        \textbf{Regime} & \textbf{Behavior} & \textbf{Interpretation}  \\
        \\
        \hline
       $\gamma \gg 1$, weak $\xi_0$ & Stable growth & Rapid actin decay; weak traction. \\          \\
       Moderate $\gamma$, high $\alpha,\ \beta$ & Oscillations & Feedback loops dominate decay, producing damped oscillations.   \\        \\
        High $\nu$, low $k$ & Instability & Fast polymerization with a weak clutch leads to runaway extension.     \\      \\
        Tuned $\nu$, $\alpha$, $\beta$, $k$ & Hopf bifurcation & Limit cycle: sustained oscillations of tip growth speed. \\
        \hline\hline
    \end{tabular}

    \caption{Regimes of growth-cone dynamics.} 
    \label{tab:I}
\end{table*}

\section{Linear Stability and Oscillations in a Stochastic Actin-Myosin-Clutch Model}

The preceding sections established a drift--diffusion description for growth cone motion using Langevin dynamics and the associated FPE, with drift fields encoding guidance and feedback, and diffusion terms capturing intrinsic and extrinsic noise. We now build on this framework by coupling the growth cone kinematics to a minimal mechanochemical model for actin polymerization, myosin--II contractility, and point--contact adhesions. This model links the probabilistic description of axon position to explicit intracellular processes, allowing us to analyze when feedback produces steady outgrowth, damped oscillations, or sustained limit cycles.

Experimental studies have established that cell-substrate adhesion is mediated by cell adhesion molecules (CAMs) which act as “molecular clutches” that couple the actin filaments (F-actin) to the underlying substrate. Growth cones form complex point contact adhesions, which are hierarchical assemblies where integrin molecules bind to extracellular matrix proteins on the growth substrate \cite{Arimura2007, Huber2003, Lowery2009}. Simultaneously, numerous other CAMs, including those involved in signal transduction and actin binding, are recruited to the intracellular side \cite{Lowery2009, Azevedo2009, Franze2010}. According to the clutch model, myosin contracts actin filaments and generates active contractile stress, while adhesion complexes produce frictional adhesion forces opposing the movement of actin filaments \cite{Franze2010, Riveline2001, Fivaz2007}.
We assume that the growth velocity $v_g$ depends on three coarse--grained internal variables: polymerized actin density $\rho_a(x,t)$, bound myosin density $M(x,t)$, and point--contact adhesion density $A(x,t)$. 

Assuming that the contractile stress is proportional to the density $M$ of myosin bound to F-actin, we use the following force balance equation for the local velocity of the actin flow in the lab coordinate system \cite{Rubinstein2009}:
\begin{equation} \nabla \cdot \left[ \mu_A \nabla v_{\text{actin}} + k  M \right] = \xi(A, \rho_a) v_{\text{actin}}, \label{balance} \end{equation}   
The first term in Equation ~\eqref{balance} is the sum of the divergences of the passive shear and deformation stresses for F-actin (where $\mu_A$ denotes the actin viscosity). The term $kM$ represents the active contractile force (with $k$ the force per myosin unit), and $\xi\!\left(A,\rho_N\right)\, v_{\textrm{actin}}$ is the frictional adhesion force. The coefficient $\xi(A, \rho_a)$ measures the strength of growth-cone--substrate adhesion and depends on the density $A$ of point-contact adhesions and on the actin density $\rho_a(x,t)$. The exact functional dependence has not been characterized in the literature. Following earlier work on keratocytes \cite{Barnhart2011}, we model adhesion with a power law dependence: $\xi(A, \rho_a) = \xi_0 A^\alpha \rho_a^\beta$, where $\xi_0$ is the adhesion coefficient and $\alpha$ and $\beta$ are feedback exponents.

Consistent with the clutch models, the actin polymerization rate scales linearly with actin density, $v_p = \nu \rho_a$. The net growth cone velocity is then given by \cite{Barnhart2011, Rubinstein2009}:
\begin{equation}
v_g = \frac{\nu \rho_a}{1 + \frac{\xi_0 A^\alpha \rho_a^\beta}{k}} -\frac{kM}{\xi_0 A^{\alpha}\rho_a^{\beta}} , 
\end{equation}
and the Fokker-Planck Equation \eqref{eq:A8} yields a Gaussian-like steady-state velocity distribution: \begin{equation} P_{\text{s}}(v) \propto \exp\left[-\frac{(v - \langle v_g \rangle)^2}{2\sigma_v^2}\right], \end{equation} where $\sigma_v^2 \sim D_g$ (effective diffusion coefficient) captures the stochastic fluctuations.

To perform the stability analysis, we adopt a coarse grained (spatially averaged) description for $(\rho_a,M,A)$ and retain the dominant kinetic couplings. Actin, myosin, and point contact dynamics are then described by \cite{Barnhart2011, Rubinstein2009}:
\begin{equation}
\frac{d\rho_a}{dt}= -\gamma \rho_a - v_g \frac{\partial \rho_a}{\partial x} \label{eq:rho}
\end{equation}

\begin{equation}
  \frac{dM}{dt} = -k_{\text{off}} M + k_{\text{on}} (M_0 - M) \label{eq:myo}\
\end{equation}

\begin{equation}
\frac{dA}{dt} = \alpha_A - \beta_A A \label{eq:adh} 
\end{equation}
where $\gamma$ is the rate of actin disassembly in the growth cone; $k_{\textrm{off}},\ k_{\textrm{on}}$ denote the myosin detachment and attachment rates, respectively; and $M_0$ is the concentration of unbound myosin (which attaches to F-actin). We also assume that point contact adhesion complexes appear across the growth cone with constant rate $\alpha_A$ and disassemble with rate $\beta_A$. 

For stability analysis we linearize Equations~\eqref{eq:rho}--\eqref{eq:adh} around the steady state $(\rho_a^*, M^*, A^*)$. 
Defining the perturbations:
\begin{equation}
\delta\rho_a = \rho_a - \rho_a^*, \quad \delta M = M - M^*, \quad \delta A = A - A^*
\end{equation}
the system ~\eqref{eq:rho}--\eqref{eq:adh} can be written as: \begin{equation} \frac{d}{dt} \begin{bmatrix} \delta \rho_a \ \delta M \ \delta A \end{bmatrix} = \mathcal{J} \begin{bmatrix} \delta \rho_a \ \delta M \ \delta A \end{bmatrix}, \end{equation} where $\mathcal{J}$ is the Jacobian matrix.  These terms form a feedback loop that quantifies how changes in each internal variable modulate the axonal drift velocity $v_g$. The local dynamics are determined by the eigenvalues of $\mathcal{J}$, which satisfy a cubic characteristic equation:
\begin{equation}
\lambda^{3}+a_{1}\lambda^{2}+a_{2}\lambda+a_{3}=0,
\label{eq:char}
\end{equation}
Routh--Hurwitz stability criterion imply linear stability when $a_{1}>0$, $a_{2}>0$, $a_{3}>0$, and $a_{1}a_{2}>a_{3}$ \cite{Murray1993, strogatz2015nonlinear}. A Hopf bifurcation (the onset of small, self--sustained oscillations) occurs when $a_{1}>0$, $a_{2}>0$, $a_{3}>0$ and $a_{1}a_{2}=a_{3}$. In the present model, oscillations arise when clutch reinforcement (large $\xi(A, \rho_a )$), load--sensitive myosin recruitment (intermediate $k_{\textrm{on}}$), or strong adhesion/actin feedback exponents ($\alpha,\beta$) amplify the effective gain in the $\rho_a\to v_g \to M,A\to v_g$ loop, so that the trace of $\mathcal{J}$ remains negative but $a_{1}a_{2}$ approaches $a_{3}$. Conversely, rapid depolymerization ($\gamma$ large), fast adhesion turnover ($\beta_A$ large), or weak mechanosensitive feedback (small $\xi_0$) push the system deep into the stable regime with purely real, negative eigenvalues (overdamped return to steady growth). A summary of the dynamical behavior is provided in Table~\ref{tab:I}. 

\begin{figure*}[!t]
    \includegraphics[width=14.5 cm]{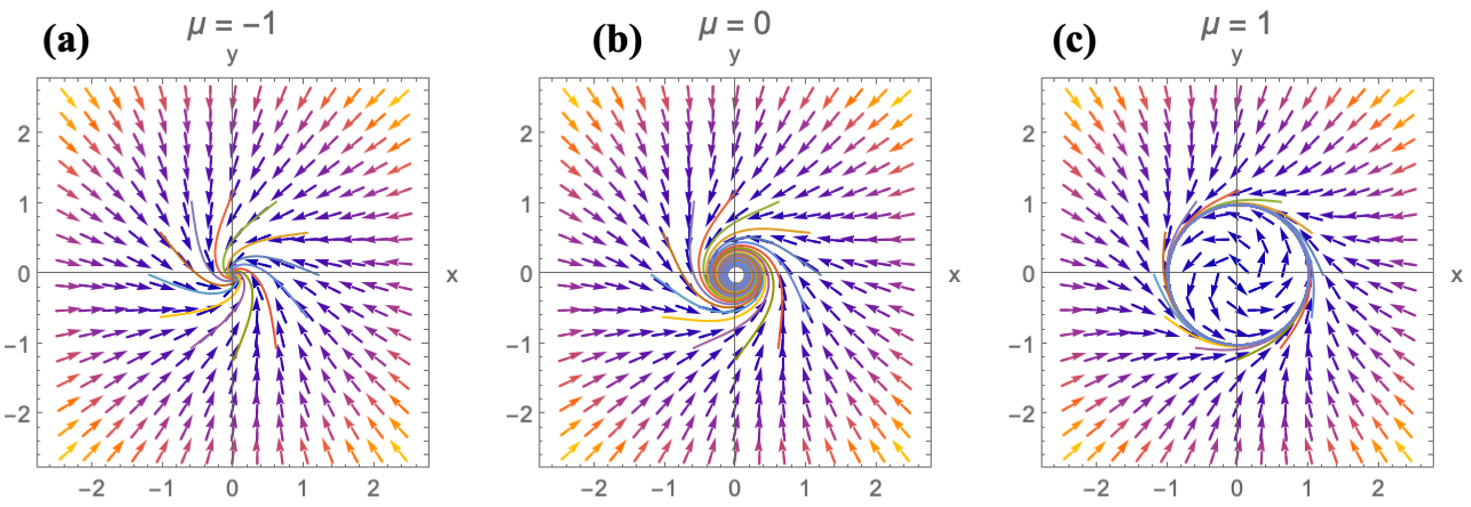}
    \caption{Phase portraits of the reduced mechanochemical actin--myosin--clutch model illustrating the transition from stable to oscillatory to unstable regimes as the feedback parameter $\mu$ is varied. 
Here, $x$ denotes the dimensionless actin-driven protrusion (tip) velocity and $y$ the normalized density of point--contact adhesions (integrins bound to the substrate), while $\mu$ represents the effective strength of feedback coupling polymerization, contractility, and adhesion reinforcement (defined in the main text).  
Trajectories evolve in the $(x,y)$ phase plane from initial conditions on a circle of radius 1.2. 
\mbox{(\textbf{a}) For $\mu=-1$}, the system exhibits stable growth (focus), with trajectories spiraling inward toward a fixed point, representing steady extension with stabilized adhesions. 
(\textbf{b}) At $\mu=0$, a Hopf bifurcation occurs and trajectories approach a closed orbit, corresponding to sustained oscillations of growth speed and clutch engagement. 
(\textbf{c}) For $\mu=1$, the origin becomes unstable and trajectories spiral outward, reflecting runaway protrusion due to dominant polymerization or weakened clutch coupling. These regimes correspond to experimentally observed transitions between steady extension, oscillatory behavior, and instability in axonal tip dynamics \cite{Betz2006, Vensi2019, Descoteaux2022, Kumarasinghe2022}. Biologically, these three regimes capture growth cones that extend steadily with stable adhesions, undergo rhythmic protrusion--retraction cycles, or experience unstable, runaway extension when feedback is \mbox{strongly amplifying.}  \label{fig3}}
\end{figure*}  

\begin{figure*}[!t]
    \includegraphics[width=13 cm]{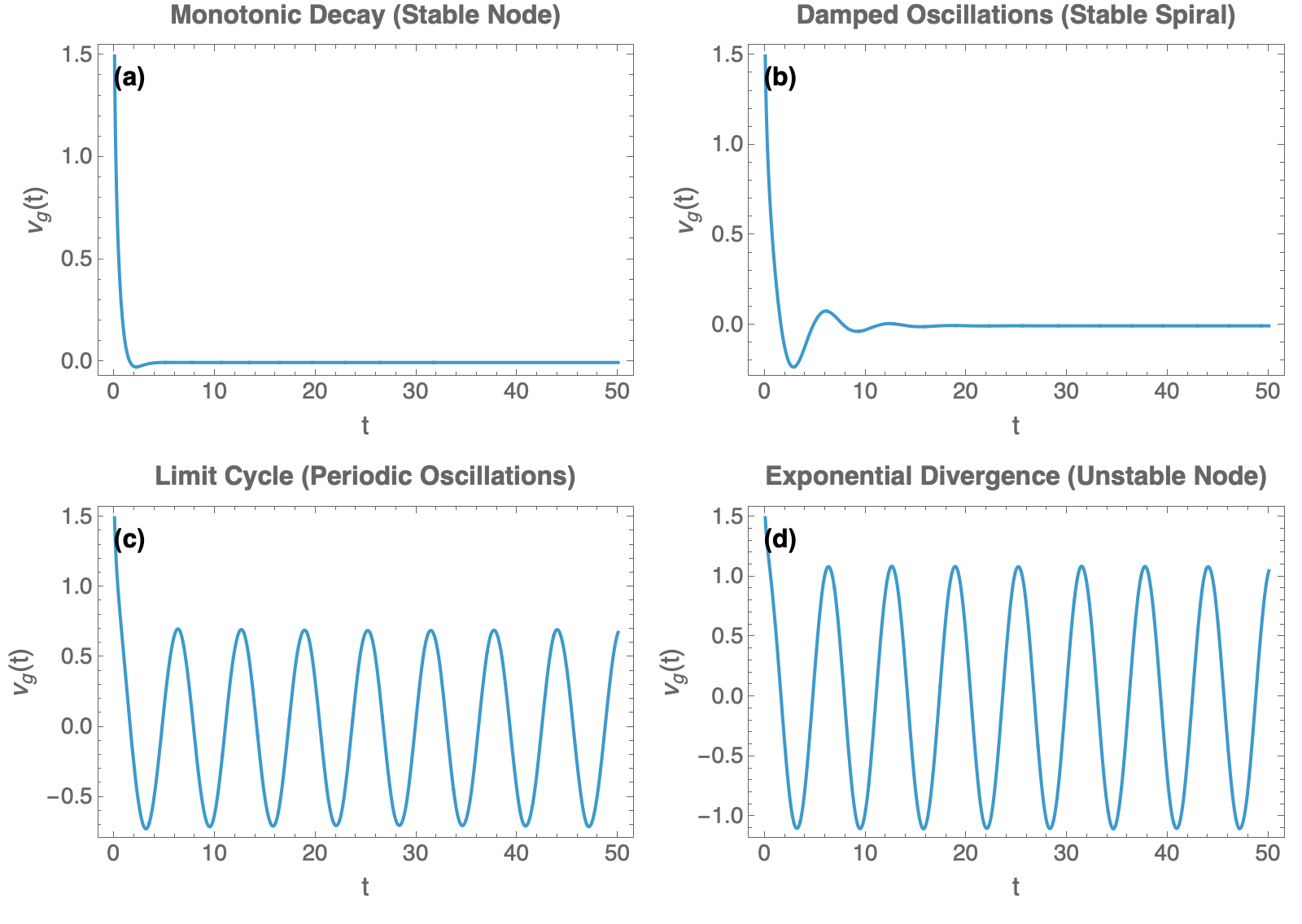}
    \caption{Time evolution of the dimensionless growth cone velocity $v_g(t)$ in the reduced mechanochemical model for different regimes of the control parameter $\mu = \operatorname{Tr}(\mathcal{J})$. The growth cone velocity $v_g(t)$ represents actin-driven protrusion modulated by nonlinear feedback between myosin contraction, adhesion-mediated resistance, and polymerization dynamics. Each panel corresponds to a different value of $\mu$, illustrating characteristic behaviors: (\textbf{a}) monotonic decay to a steady state for $\mu = -1.5$ (stable node); (\textbf{b}) damped oscillations for $\mu = -0.3$ (stable spiral); (\textbf{c}) sustained periodic oscillations (limit cycle) for $\mu = 0.05$; (\textbf{d}) exponential divergence for $\mu = 1.2$ (unstable node). In panels (\textbf{c},\textbf{d}), the oscillation crosses $v_g = 0$ and reaches values down to $v_g \approx -1$. Negative values indicate short episodes of net growth cone retraction (backward motion) when contractile forces and adhesion-mediated friction transiently dominate polymerization-driven advance.  \label{fig4}}
\end{figure*}   

\textit{Illustrative eigenvalue spectrum.} As a concrete example, we consider a reduced mechanochemical model of growth cone dynamics. Let \(x\) denote the dimensionless, actin-driven protrusion (tip) velocity, normalized by a characteristic polymerization speed. This quantity captures contributions from actin polymerization, myosin contractility, and cytoskeletal resistance. Let \(y\) denote the normalized adhesion density, proportional to the number of substrate-bound point--contact adhesions (e.g., integrin complexes), and scaled by a reference density. The pair (\(x\), \(y\)) constitutes the minimal state vector for modeling the coupled dynamics of axonal extension and adhesion regulation under feedback control.
For the dynamical variables \mbox{\(x\)  and \(y\),} the Jacobian is
\begin{equation}
J =
\begin{pmatrix}
\frac{\partial \dot{x}}{\partial x} & \frac{\partial \dot{x}}{\partial y} \\
\frac{\partial \dot{y}}{\partial x} & \frac{\partial \dot{y}}{\partial y}
\end{pmatrix}_{(x^*,y^*)},
\end{equation}
and we define the effective bifurcation parameter \(\mu\) as the trace of the Jacobian matrix of the linearized system at the steady state $(x^*,y^*)$ \cite{strogatz2015nonlinear, Murray1993}:
\begin{equation}
 \mu \;=\; \mathrm{Tr}\,J \;=\;
\left.\frac{\partial \dot{x}}{\partial x}\right|_{(x^*,y^*)}+
\left.\frac{\partial \dot{y}}{\partial y}\right|_{(x^*,y^*)}.
\end{equation}

In the reduced mechanochemical model of actin-driven protrusion and clutch-mediated force transmission, the dimensionless bifurcation parameter \(\mu\) can be interpreted as a composite control parameter encoding the balance between actin polymerization, substrate adhesion, and myosin-driven contractility \cite{Murray1993, Cheng2025}. Specifically, \(\mu\) increases with higher actin polymerization rate, which drives forward motion; stronger effective adhesion \(\xi(A,\rho)\), which promotes traction force buildup; and reduced myosin contractility \(k \nabla M\), which otherwise resists forward motion. Thus, \(\mu\) acts as an effective propulsion-to-friction ratio. The phase portraits for the reduced mechaniochemical model are shown in \mbox{(Figure~\ref{fig3}).} Negative values of \(\mu\) correspond to a stable focus with damped protrusion dynamics \mbox{(Figure~\ref{fig3}a);} \(\mu=0\) marks a Hopf bifurcation (Figure~\ref{fig3}b); and positive \(\mu\) indicates an unstable focus or limit-cycle oscillations (Figure~\ref{fig3}c).

Figure~\ref{fig4} shows the corresponding temporal profiles of the dimensionless growth cone velocity \(v_g(t)\) for the same regimes of the control parameter $\mu$ as in Figure~\ref{fig3}. \mbox{For \(\mu<0\), trajectories} that spiral toward the fixed point in Figure~\ref{fig3}a manifest as monotonic relaxation or damped oscillations in time (Figure~\ref{fig4}a,b). Near the Hopf threshold (\(\mu\approx 0\)), the approach to a closed orbit in Figure~\ref{fig3}b becomes a stable limit cycle with sustained oscillations of \(v_g(t)\) (Figure~\ref{fig4}c). For \(\mu>0\), the outward spirals or repelling fixed points in Figure~\ref{fig3}c are reflected by exponential growth of \(v_g(t)\) (Figure~\ref{fig4}d). Together, these time traces confirm the bifurcation structure inferred from the eigenvalue analysis and provide observable readouts that distinguish stable, weakly underdamped, limit-cycle, and \mbox{unstable regimes.}

\section{Discussion and Conclusions}

The results presented here connect a stochastic description of axonal motion with an explicit mechanochemical account of growth cone regulation. In the first part of the paper, Langevin dynamics and the associated Fokker--Planck equation (FPE) provide \mbox{a drift--diffusion} framework for axonal trajectories, with drift fields encoding guidance and feedback and diffusion terms capturing intrinsic and extrinsic fluctuations. We then coupled this FPE drift to a minimal actin--myosin--clutch model that links polymerization-driven protrusion, myosin-II contractility, and point--contact adhesion. This integration yields a compact model in which measurable intracellular processes set the parameters of \mbox{a stochastic} transport equation, enabling direct comparison to trajectory statistics.

The linear stability analysis clarifies how positive and negative feedback loops jointly regulate robust yet adaptable outgrowth. Near a stable fixed point, linearizing the FPE with approximately constant effective diffusivity reduces the growth cone dynamics to \mbox{an Ornstein--Uhlenbeck} process. This predicts a Gaussian stationary distribution of growth cone velocities narrowly peaked around $\langle v_g \rangle$ with variance controlled by effective diffusion coefficient $D_g$. As parameters approach a Hopf bifurcation, the dominant eigenvalues of the coarse grained mechanochemical dynamics form a weakly damped complex pair. In this regime, stochastic trajectories exhibit quasi-oscillatory fluctuations, providing concrete experimental signatures of clutch-mediated feedback. Beyond the Hopf threshold, the model admits a stable limit cycle in the coarse-grained variables, consistent with self-sustained growth--retraction oscillations whose amplitude and frequency are set by the internal feedbacks rather than by initial conditions.

Biologically, these regimes map onto experimentally accessible control parameters. Strengthening adhesion reinforcement or increasing load-sensitive myosin recruitment raises the closed-loop gain, pushing the system toward oscillatory dynamics. In contrast, rapid actin turnover or fast adhesion disassembly lowers the gain and stabilizes steady extension. At the level of substrate mechanics and adhesion, increasing anisotropy or curvature in a way that promotes alignment tends to strengthen the drift and reduce angular diffusion, whereas softening the substrate or weakening integrin engagement increases the effective diffusion and can suppress oscillations by limiting traction forces. In this framework, coupled positive and negative feedbacks generate reliable morphogenic outcomes: positive feedback amplifies weak guidance cues into directed motion, while negative feedback constrains fluctuations and allows rapid adaptation to changes in the biochemical and mechanical environment. 

The model yields several testable predictions that link intracellular control to ensemble statistics. In the stable regime, the Ornstein--Uhlenbeck reduction implies exponentially decaying velocity autocorrelations with a single characteristic time and a Gaussian-like distribution of short-time displacements \cite{Staii2024}. As the system approaches a Hopf bifurcation, the dynamics becomes oscillatory and the power spectrum acquires a narrow peak that sharpens as damping decreases. Pharmacological or genetic perturbations provide targeted validation: partial inhibition of myosin-II (reducing the contractile coefficient) should shift the spectrum toward lower frequency and diminish oscillation amplitude; interventions that enhance adhesion reinforcement (e.g., integrin activation) should have the opposite effect. The parameters required by the model---including effective motility coefficients, correlation times, and measures of cell--substrate coupling---can be estimated by combining trajectory analysis on patterned substrates with independent mechanical measurements (for example, atomic force and traction force microscopy measurements \cite{Spedden2012Elasticity, Kumarasinghe2022}), enabling \mbox{a quantitative} model-experiment comparison.

The present framework is closely related to, but distinct from, other widely used descriptions of cell and growth cone motility. Biased persistent random walk models and velocity-jump processes treat trajectories as piecewise constant motions punctuated by reorientation events. In the small-step limit, these converge to drift--diffusion equations with effective bias and persistence lengths, and our Langevin/Fokker--Planck formulation can be viewed as a continuous-time counterpart in which the bias and angular diffusion are explicitly linked to substrate geometry and feedback-controlled clutch dynamics. Likewise, the Ornstein--Uhlenbeck velocity models commonly used in cell motility are recovered as a special case of our framework for motion on uniform substrates with linear drift and constant diffusion. Reaction--diffusion sensing models for growth cone guidance, by contrast, emphasize spatiotemporal ligand profiles and intracellular signaling rather than mechanics. The present approach is complementary in that it treats these cues as inputs that modulate the drift and diffusion terms, allowing mechanical and chemical guidance to be incorporated on equal footing in a single stochastic transport description.

An important implication of this analysis is that the model makes predictions that are not only experimentally testable but also quantitatively distinguishable from alternative modeling approaches. In drift--diffusion or biased persistent random walk models without explicit mechanochemical feedback, both the mean growth cone velocity and the velocity autocorrelation function typically relax monotonically to a steady state, with perturbations decaying without oscillations. By contrast, when clutch-mediated feedback drives the system close to a Hopf bifurcation, our framework predicts weakly damped or sustained oscillations in the coarse-grained growth cone velocity, with the characteristic period and degree of damping determined by the adhesion feedback exponents, myosin on/off rates, and the effective polymerization speed. Similarly, the stationary angular distributions exhibit characteristic sharpening as the deterministic torque is varied, and the crossover from diffusive to superdiffusive mean-square displacement on micropatterned surfaces provides an additional quantitative signature for distinguishing between models. Systematic measurements of these observables, combined with independent mechanical assays of traction forces and cell stiffness, therefore offer a concrete strategy to validate the model and to distinguish clutch-regulated stochastic dynamics from purely kinematic or purely reaction--diffusion descriptions of growth cone guidance.

Several limitations suggest directions for refinement. The coarse-grained mechanochemical variables capture dominant feedbacks but neglect spatial heterogeneity across the growth cone, time delays in signaling, and curvature-dependent kinematics. Extending the internal dynamics to spatially distributed fields would capture competition between local protrusion and retrograde actin flow and enable mode-selection analyses of spatiotemporal patterns. Likewise, the present FPE treats drift and diffusion as functions of coarse variables. When diffusion depends on position or orientation (multiplicative noise), the stochastic calculus convention (Itô versus Stratonovich) and any induced noise-interpretation drift (spurious-drift terms) should be specified and tested. At the population level, coupling single-axon dynamics through interaction rules (fasciculation/defasciculation, and competition for cues) would bridge single-cell biophysics to the formation of axon bundles and neuronal network architecture. Finally, parameter identifiability warrants careful analysis: joint fits to displacement distributions, autocorrelations, and power spectra, together with independent mechanical measurements, provide complementary constraints that improve robustness and reduce model degeneracy.

In conclusion, by embedding a clutch-regulated mechanochemical model into the drift term of the Fokker--Planck equation, we obtain a predictive, general framework that links intracellular regulation to stochastic growth cone motion and observed trajectory statistics. The stability analysis identifies biologically meaningful regimes: steady extension, damped oscillations, and sustained limit cycles, and specifies how polymerization, contractility, and adhesion feedback tune transitions among them. Beyond clarifying mechanisms of axonal guidance and connectivity, this framework provides practical design rules for engineered substrates and neuroprosthetic scaffolds: pattern anisotropy and curvature modulate directional drift; adhesion chemistry and stiffness tune diffusion and closed-loop gain; and targeted perturbations shift the system toward or away from oscillatory regimes associated with exploratory growth. From an applied-mathematics standpoint, the framework unifies SDE/FPE modeling with linear and nonlinear stability analysis, and provides concrete routes for parameter estimation and control-oriented design. Together, these insights advance a quantitative framework for controlling axonal outgrowth in vitro and, ultimately, for promoting functional regeneration in engineered microenvironments.

\section*{acknowledgments}
This research was funded by a Tufts Faculty Research Award (FRAC).

\bibliography{NewCitation}

@article{Staii2024,
  author       = {Staii, Cristian},
  title        = {Nonlinear Growth Dynamics of Neuronal Cells Cultured on Directional Surfaces},
  journal      = {Biomimetics},
  year         = {2024},
  volume       = {9},
  number       = {4},
  pages        = {203},
  issn         = {2313-7673},
  url          = {https://www.mdpi.com/2313-7673/9/4/203},
  doi          = {10.3390/biomimetics9040203},
  publisher    = {MDPI},
  month        = {April},
  type         = {Journal Article}
}

@article{Staii2023,
   author = {Staii, C.},
   title = {Biased Random Walk Model of Neuronal Dynamics on Substrates with Periodic Geometrical Patterns},
   journal = {Biomimetics (Basel)},
   volume = {8},
   number = {2},
   ISSN = {2313-7673},
   DOI = {10.3390/biomimetics8020267},
   year = {2023},
   type = {Journal Article}
}

@article{CastroDominguez2023,
   author = {Castro-Domínguez, C. and Lozano-Picazo, P. and Álvarez-López, A. and Garrote-Junco, J. and Panetsos, F. and Guinea, G.V. and Elices, M. and Rojo, F.J. and González-Nieto, D. and Colchero, L. and others},
   title = {Axonal Guidance Using Biofunctionalized Straining Flow Spinning Regenerated Silk Fibroin Fibers as Scaffold},
   journal = {Biomimetics},
   volume = {8},
   number = {1},
   pages = {65},
   year = {2023},
   doi = {10.3390/biomimetics8010065},
   issn = {2313-7673},
   url = {https://www.mdpi.com/2313-7673/8/1/65},
   type = {Journal Article}
}

@article{Oliveri2022,
   author = {Oliveri, H. and Goriely, A.},
   title = {Mathematical models of neuronal growth},
   journal = {Biomech Model Mechanobiol},
   volume = {21},
   number = {1},
   pages = {89-118},
   ISSN = {1617-7959 (Print)
1617-7940},
   DOI = {10.1007/s10237-021-01539-0},
   year = {2022},
   type = {Journal Article}
}

@article{Kumarasinghe2022,
   author = {Kumarasinghe, U. and Fox, L. N. and Staii, C.},
   title = {Combined Traction Force-Atomic Force Microscopy Measurements of Neuronal Cells},
   journal = {Biomimetics (Basel)},
   volume = {7},
   number = {4},
   ISSN = {2313-7673},
   DOI = {10.3390/biomimetics7040157},
   year = {2022},
   type = {Journal Article}
}

@article{Descoteaux2022,
   author = {Descoteaux, M. and Sunnerberg, J. P. and Brady, D. D. and Staii, C.},
   title = {Feedback-controlled dynamics of neuronal cells on directional surfaces},
   journal = {Biophys J},
   volume = {121},
   number = {5},
   pages = {769-781},
   ISSN = {0006-3495 (Print)
0006-3495},
   DOI = {10.1016/j.bpj.2022.01.020},
   year = {2022},
   type = {Journal Article}
}

@article{Yurchenko2021,
   author = {Yurchenko, I. and Farwell, M. and Brady, D. D. and Staii, C.},
   title = {Neuronal Growth and Formation of Neuron Networks on Directional Surfaces},
   journal = {Biomimetics (Basel)},
   volume = {6},
   number = {2},
   ISSN = {2313-7673},
   DOI = {10.3390/biomimetics6020041},
   year = {2021},
   type = {Journal Article}
}

@article{El-Samad2021,
   author = {El-Samad, H.},
   title = {Biological feedback control-Respect the loops},
   journal = {Cell Syst},
   volume = {12},
   number = {6},
   pages = {477-487},
   ISSN = {2405-4712},
   DOI = {10.1016/j.cels.2021.05.004},
   year = {2021},
   type = {Journal Article}
}

@article{Teixeira2020,
   author = {Teixeira, M. I. and Amaral, M. H. and Costa, P. C. and Lopes, C. M. and Lamprou, D. A.},
   title = {Recent Developments in Microfluidic Technologies for Central Nervous System Targeted Studies},
   journal = {Pharmaceutics},
   volume = {12},
   number = {6},
   ISSN = {1999-4923 (Print)
1999-4923},
   DOI = {10.3390/pharmaceutics12060542},
   year = {2020},
   type = {Journal Article}
}

@article{Buskermolen2020,
   author = {Buskermolen, A. B. C. and Ristori, T. and Mostert, D. and van Turnhout, M. C. and Shishvan, S. S. and Loerakker, S. and Kurniawan, N. A. and Deshpande, V. S. and Bouten, C. V. C.},
   title = {Cellular Contact Guidance Emerges from Gap Avoidance},
   journal = {Cell Rep Phys Sci},
   volume = {1},
   number = {5},
   pages = {100055},
   ISSN = {2666-3864},
   DOI = {10.1016/j.xcrp.2020.100055},
   year = {2020},
   type = {Journal Article}
}

@article{Alert2020,
   author = {Alert, Ricard and Trepat, Xavier},
   title = {Physical models of collective cell migration},
   journal = {Annual Review of Condensed Matter Physics},
   volume = {11},
   number = {1},
   pages = {77-101},
   ISSN = {1947-5454},
   year = {2020},
   type = {Journal Article}
}

@article{Yurchenko2019,
   author = {Yurchenko, I. and Vensi Basso, J. M. and Syrotenko, V. S. and Staii, C.},
   title = {Anomalous diffusion for neuronal growth on surfaces with controlled geometries},
   journal = {PLoS One},
   volume = {14},
   number = {5},
   pages = {e0216181},
   ISSN = {1932-6203},
   DOI = {10.1371/journal.pone.0216181},
   year = {2019},
   type = {Journal Article}
}

@article{Vensi2019,
   author = {Vensi Basso, Joao Marcos and Yurchenko, Ilya and Simon, Marc and Rizzo, Daniel J. and Staii, Cristian},
   title = {Role of geometrical cues in neuronal growth},
   journal = {Physical Review E},
   volume = {99},
   number = {2},
   pages = {022408},
   note = {PRE},
   DOI = {10.1103/PhysRevE.99.022408},
   url = {https://link.aps.org/doi/10.1103/PhysRevE.99.022408},
   year = {2019},
   type = {Journal Article}
}

@article{Takano2019,
   author = {Takano, T. and Funahashi, Y. and Kaibuchi, K.},
   title = {Neuronal Polarity: Positive and Negative Feedback Signals},
   journal = {Front Cell Dev Biol},
   volume = {7},
   pages = {69},
   ISSN = {2296-634X (Print)
2296-634x},
   DOI = {10.3389/fcell.2019.00069},
   year = {2019},
   type = {Journal Article}
}

@article{Basso2019,
   author = {Basso, Joao Marcos Vensi and Yurchenko, Ilya and Wiens, Matthew R. and Staii, Cristian},
   title = {Neuron dynamics on directional surfaces},
   journal = {Soft Matter},
   volume = {15},
   number = {48},
   pages = {9931-9941},
   ISSN = {1744-683X},
   DOI = {10.1039/C9SM01769K},
   url = {http://dx.doi.org/10.1039/C9SM01769K},
   year = {2019},
   type = {Journal Article}
}

@article{Lilja2018,
   author = {Lilja, J. and Ivaska, J.},
   title = {Integrin activity in neuronal connectivity},
   journal = {J Cell Sci},
   volume = {131},
   number = {12},
   ISSN = {0021-9533},
   DOI = {10.1242/jcs.212803},
   year = {2018},
   type = {Journal Article}
}

@article{Polackwich2015,
   author = {Polackwich, R. J. and Koch, D. and McAllister, R. and Geller, H. M. and Urbach, J. S.},
   title = {Traction force and tension fluctuations in growing axons},
   journal = {Front Cell Neurosci},
   volume = {9},
   pages = {417},
   ISSN = {1662-5102 (Print)
1662-5102},
   DOI = {10.3389/fncel.2015.00417},
   year = {2015},
   type = {Journal Article}
}

@article{Jurchenko2015,
   author = {Jurchenko, C. and Salaita, K. S.},
   title = {Lighting Up the Force: Investigating Mechanisms of Mechanotransduction Using Fluorescent Tension Probes},
   journal = {Mol Cell Biol},
   volume = {35},
   number = {15},
   pages = {2570-82},
   ISSN = {0270-7306 (Print)
0270-7306},
   DOI = {10.1128/mcb.00195-15},
   year = {2015},
   type = {Journal Article}
}

@article{Moeendarbary2014,
   author = {Moeendarbary, E. and Harris, A. R.},
   title = {Cell mechanics: principles, practices, and prospects},
   journal = {Wiley Interdiscip Rev Syst Biol Med},
   volume = {6},
   number = {5},
   pages = {371-88},
   ISSN = {1939-5094 (Print)
1939-005x},
   DOI = {10.1002/wsbm.1275},
   year = {2014},
   type = {Journal Article}
}

@article{Hyland2014,
   author = {Hyland, Callen and Mertz, Aaron F. and Forscher, Paul and Dufresne, Eric},
   title = {Dynamic peripheral traction forces balance stable neurite tension in regenerating Aplysia bag cell neurons},
   journal = {Scientific Reports},
   volume = {4},
   number = {1},
   pages = {4961},
   ISSN = {2045-2322},
   DOI = {10.1038/srep04961},
   url = {https://doi.org/10.1038/srep04961},
   year = {2014},
   type = {Journal Article}
}

@article{Rizzo2013,
   author = {Rizzo, Daniel J. and White, James D. and Spedden, Elise and Wiens, Matthew R. and Kaplan, David L. and Atherton, Timothy J. and Staii, Cristian},
   title = {Neuronal growth as diffusion in an effective potential},
   journal = {Physical Review E},
   volume = {88},
   number = {4},
   pages = {042707},
   note = {PRE},
   DOI = {10.1103/PhysRevE.88.042707},
   url = {https://link.aps.org/doi/10.1103/PhysRevE.88.042707},
   year = {2013},
   type = {Journal Article}
}

@article{Franze2013,
   author = {Franze, K.},
   title = {The mechanical control of nervous system development},
   journal = {Development},
   volume = {140},
   number = {15},
   pages = {3069-77},
   ISSN = {0950-1991},
   DOI = {10.1242/dev.079145},
   year = {2013},
   type = {Journal Article}
}

@article{Pouwels2012,
   author = {Pouwels, J. and Nevo, J. and Pellinen, T. and Ylänne, J. and Ivaska, J.},
   title = {Negative regulators of integrin activity},
   journal = {J Cell Sci},
   volume = {125},
   number = {Pt 14},
   pages = {3271-80},
   ISSN = {0021-9533},
   DOI = {10.1242/jcs.093641},
   year = {2012},
   type = {Journal Article}
}

@article{Koch2012,
   author = {Koch, D. and Rosoff, W. J. and Jiang, J. and Geller, H. M. and Urbach, J. S.},
   title = {Strength in the periphery: growth cone biomechanics and substrate rigidity response in peripheral and central nervous system neurons},
   journal = {Biophys J},
   volume = {102},
   number = {3},
   pages = {452-60},
   ISSN = {0006-3495 (Print)
0006-3495},
   DOI = {10.1016/j.bpj.2011.12.025},
   year = {2012},
   type = {Journal Article}
}

@article{Pfister2011,
   author = {Pfister, B. J. and Gordon, T. and Loverde, J. R. and Kochar, A. S. and Mackinnon, S. E. and Cullen, D. K.},
   title = {Biomedical engineering strategies for peripheral nerve repair: surgical applications, state of the art, and future challenges},
   journal = {Crit Rev Biomed Eng},
   volume = {39},
   number = {2},
   pages = {81-124},
   ISSN = {0278-940X (Print)
0278-940x},
   DOI = {10.1615/critrevbiomedeng.v39.i2.20},
   year = {2011},
   type = {Journal Article}
}

@article{Toriyama2010,
   author = {Toriyama, M. and Sakumura, Y. and Shimada, T. and Ishii, S. and Inagaki, N.},
   title = {A diffusion-based neurite length-sensing mechanism involved in neuronal symmetry breaking},
   journal = {Mol Syst Biol},
   volume = {6},
   pages = {394},
   ISSN = {1744-4292},
   DOI = {10.1038/msb.2010.51},
   year = {2010},
   type = {Journal Article}
}

@article{Franze2010,
   author = {Franze, Kristian and Guck, Jochen},
   title = {The biophysics of neuronal growth},
   journal = {Reports on Progress in Physics},
   volume = {73},
   number = {9},
   pages = {094601},
   ISSN = {0034-4885},
   DOI = {10.1088/0034-4885/73/9/094601},
   url = {https://dx.doi.org/10.1088/0034-4885/73/9/094601},
   year = {2010},
   type = {Journal Article}
}

@article{Lowery2009,
   author = {Lowery, L. A. and Van Vactor, D.},
   title = {The trip of the tip: understanding the growth cone machinery},
   journal = {Nat Rev Mol Cell Biol},
   volume = {10},
   number = {5},
   pages = {332-43},
   ISSN = {1471-0072 (Print)
1471-0072},
   DOI = {10.1038/nrm2679},
   year = {2009},
   type = {Journal Article}
}

@article{Franze2009,
   author = {Franze, K. and Gerdelmann, J. and Weick, M. and Betz, T. and Pawlizak, S. and Lakadamyali, M. and Bayer, J. and Rillich, K. and Gögler, M. and Lu, Y. B. and Reichenbach, A. and Janmey, P. and Käs, J.},
   title = {Neurite branch retraction is caused by a threshold-dependent mechanical impact},
   journal = {Biophys J},
   volume = {97},
   number = {7},
   pages = {1883-90},
   ISSN = {0006-3495 (Print)
0006-3495},
   DOI = {10.1016/j.bpj.2009.07.033},
   year = {2009},
   type = {Journal Article}
}

@article{Fivaz2007,
   author = {Fivaz, M. and Bandara, S. and Inoue, T. and Meyer, T.},
   title = {Robust neuronal symmetry breaking by Ras-triggered local positive feedback},
   journal = {Curr Biol},
   volume = {18},
   number = {1},
   pages = {44-50},
   ISSN = {0960-9822 (Print)
0960-9822},
   DOI = {10.1016/j.cub.2007.11.051},
   year = {2008},
   type = {Journal Article}
}

@article{Arimura2007,
   author = {Arimura, N. and Kaibuchi, K.},
   title = {Neuronal polarity: from extracellular signals to intracellular mechanisms},
   journal = {Nat Rev Neurosci},
   volume = {8},
   number = {3},
   pages = {194-205},
   ISSN = {1471-003X (Print)
1471-003x},
   DOI = {10.1038/nrn2056},
   year = {2007},
   type = {Journal Article}
}

@article{Medeiros2006,
   author = {Medeiros, N. A. and Burnette, D. T. and Forscher, P.},
   title = {Myosin II functions in actin-bundle turnover in neuronal growth cones},
   journal = {Nat Cell Biol},
   volume = {8},
   number = {3},
   pages = {215-26},
   ISSN = {1465-7392 (Print)
1465-7392},
   DOI = {10.1038/ncb1367},
   year = {2006},
   type = {Journal Article}
}

@article{Huber2003,
   author = {Huber, A. B. and Kolodkin, A. L. and Ginty, D. D. and Cloutier, J. F.},
   title = {Signaling at the growth cone: ligand-receptor complexes and the control of axon growth and guidance},
   journal = {Annu Rev Neurosci},
   volume = {26},
   pages = {509-63},
   ISSN = {0147-006X (Print)
0147-006x},
   DOI = {10.1146/annurev.neuro.26.010302.081139},
   year = {2003},
   type = {Journal Article}
}

@article{Riveline2001,
   author = {Riveline, D. and Zamir, E. and Balaban, N. Q. and Schwarz, U. S. and Ishizaki, T. and Narumiya, S. and Kam, Z. and Geiger, B. and Bershadsky, A. D.},
   title = {Focal contacts as mechanosensors: externally applied local mechanical force induces growth of focal contacts by an mDia1-dependent and ROCK-independent mechanism},
   journal = {J Cell Biol},
   volume = {153},
   number = {6},
   pages = {1175-86},
   ISSN = {0021-9525 (Print)
0021-9525},
   DOI = {10.1083/jcb.153.6.1175},
   year = {2001},
   type = {Journal Article}
}

@article{Samuels1996,
   author = {Samuels, D. C. and Hentschel, H. G. and Fine, A.},
   title = {The origin of neuronal polarization: a model of axon formation},
   journal = {Philos Trans R Soc Lond B Biol Sci},
   volume = {351},
   number = {1344},
   pages = {1147-56},
   ISSN = {0962-8436 (Print)
0962-8436},
   DOI = {10.1098/rstb.1996.0099},
   year = {1996},
   type = {Journal Article}
}

@article{Collier1996,
   author = {Collier, J. R. and Monk, N. A. and Maini, P. K. and Lewis, J. H.},
   title = {Pattern formation by lateral inhibition with feedback: a mathematical model of delta-notch intercellular signalling},
   journal = {J Theor Biol},
   volume = {183},
   number = {4},
   pages = {429-46},
   ISSN = {0022-5193 (Print)
0022-5193},
   DOI = {10.1006/jtbi.1996.0233},
   year = {1996},
   type = {Journal Article}
}

@article{Ahmadzadeh2015,
   title = {Mechanical Effects of Dynamic Binding between Tau Proteins on Microtubules during Axonal Injury},
    journal = {Biophysical Journal},
    volume = {109},
    number = {11},
    pages = {2328-2337},
    year = {2015},
    issn = {0006-3495},
    doi = {https://doi.org/10.1016/j.bpj.2015.09.010},
    url = {https://www.sciencedirect.com/science/article/pii/S0006349515009443},
    author = {Hossein Ahmadzadeh and Douglas. H. Smith and Vivek B. Shenoy}
}

@article{Athamneh2017,
	author = {Athamneh, Ahmad I. M. and He, Yingpei and Lamoureux, Phillip and Fix, Lucas and Suter, Daniel M. and Miller, Kyle E.},
	date = {2017/08/04},
	doi = {10.1038/s41598-017-07402-6},
	isbn = {2045-2322},
	journal = {Scientific Reports},
	number = {1},
	pages = {7292},
	title = {Neurite elongation is highly correlated with bulk forward translocation     of microtubules},
	url = {https://doi.org/10.1038/s41598-017-07402-6},
	volume = {7},
	year = {2017}
}

@article{Azevedo2009,
  author  = {Azevedo, Frederico A. C. and Carvalho, Ludmila R. B. and Grinberg, Lea T. and Farfel, Jos{\'e} Marcelo and Ferretti, Renata E. L. and Leite, Renata E. P. and Filho, Wilson Jacob and Lent, Roberto and Herculano-Houzel, Suzana},
  title   = {Equal numbers of neuronal and nonneuronal cells make the human brain an isometrically scaled-up primate brain},
  journal = {Journal of Comparative Neurology},
  volume  = {513},
  number  = {5},
  pages   = {532--541},
  year    = {2009},
  doi     = {10.1002/cne.21974}
}

@article{Bayly2014,
    title = {Mechanical forces in cerebral cortical folding: A review of measurements and models},
    journal = {Journal of the Mechanical Behavior of Biomedical Materials},
    volume = {29},
    pages = {568-581},
    year = {2014},
    issn = {1751-6161},
    doi = {https://doi.org/10.1016/j.jmbbm.2013.02.018},
    url = {https://www.sciencedirect.com/science/article/pii/S175161611300074X},
    author = {P.V. Bayly and L.A. Taber and C.D. Kroenke}
}

@article{DeGennes2007,
  author  = {de Gennes, P.-G.},
  title   = {Collective neuronal growth and self-organization of axons},
  journal = {Proceedings of the National Academy of Sciences},
  volume  = {104},
  number  = {12},
  pages   = {4904--4906},
  year    = {2007},
  doi     = {10.1073/pnas.0609871104}
}

@article{Katz1984,
  title   = {Axonal Elongation as a Stochastic Walk},
  author  = {Katz, M. J. and George, E. B. and Gilbert, L. J.},
  journal = {Cell Motil.},
  volume  = {4},
  pages   = {351--370},
  year    = {1984},
  doi     = {10.1002/cm.970040505},
  url     = {https://doi.org/10.1002/cm.970040505}
}

@article{DeRooij2018a,
    title = {Microtubule Polymerization and Cross-Link Dynamics Explain Axonal Stiffness and Damage},
    journal = {Biophysical Journal},
    volume = {114},
    number = {1},
    pages = {201-212},
    year = {2018},
    issn = {0006-3495},
    doi = {https://doi.org/10.1016/j.bpj.2017.11.010},
    url = {https://www.sciencedirect.com/science/article/pii/S0006349517312390},
    author = {Rijk {de Rooij} and Ellen Kuhl}
}

@ARTICLE{DeRooij2018b,
    author = {de Rooij, Rijk  and Kuhl, Ellen },
    title = {Physical Biology of Axonal Damage},
    journal = {Frontiers in Cellular Neuroscience},
    volume = {12},
    year = {2018},
    URL={https://www.frontiersin.org/journals/cellular-neuroscience/articles/10.3389/fncel.2018.00144},
    doi = {10.3389/fncel.2018.00144},
    issn = {1662-5102}
}

@article{Franze2020,
   author = "Franze, Kristian",
   title = "Integrating Chemistry and Mechanics: The Forces Driving Axon Growth", 
   journal= "Annual Review of Cell and Developmental Biology",
   year = "2020",
   volume = "36",
   number = "Volume 36, 2020",
   pages = "61-83",
   doi = "https://doi.org/10.1146/annurev-cellbio-100818-125157",
   url = "https://www.annualreviews.org/content/journals/10.1146/annurev-cellbio-100818-125157",
   publisher = "Annual Reviews",
   issn = "1530-8995",
   type = "Journal Article",
}

@article{Goodhill1999,
  author  = {Goodhill, Geoffrey J. and Urbach, Jeffrey S.},
  title   = {Theoretical analysis of gradient detection by growth cones},
  journal = {Journal of Neurobiology},
  volume  = {41},
  number  = {2},
  pages   = {230--241},
  year    = {1999},
  doi     = {10.1002/(SICI)1097-4695(19991105)41:2<230::AID-NEU6>3.0.CO;2-9}
}

@inbook{Goodhill2003,
    author = {Goodhill, Geoffrey and Urbach, Jeffrey},
    year = {2003},
    month = {08},
    pages = {95-110},
    title = {Axon Guidance and Gradient Detection by Growth Cones},
    isbn = {9780262285438},
    doi = {10.7551/mitpress/4703.003.0007},
    publisher = {The MIT Press, Development, Cambridge}
}

@article{Holland2015,
    author = {Holland, Maria A. and Miller, Kyle E. and Kuhl, Ellen},
    date = {2015/07/01},
    doi = {10.1007/s10439-015-1312-9},
    id = {Holland2015},
    isbn = {1573-9686},
    journal = {Annals of Biomedical Engineering},
    number = {7},
    pages = {1640--1653},
    title = {Emerging Brain Morphologies from Axonal Elongation},
    url = {https://doi.org/10.1007/s10439-015-1312-9},
    volume = {43},
    year = {2015}
}

@article{Jakobs2020,
    title = {Mechanical Regulation of Neurite Polarization and Growth: A Computational Study},
    journal = {Biophysical Journal},
    volume = {118},
    number = {8},
    pages = {1914-1920},
    year = {2020},
    issn = {0006-3495},
    doi = {https://doi.org/10.1016/j.bpj.2020.02.031},
    url = {https://www.sciencedirect.com/science/article/pii/S0006349520302356},
    author = {Maximilian A.H. Jakobs and Kristian Franze and Assaf Zemel}
}

@article{Kiryushko2004,
  author  = {Kiryushko, Darya and Berezin, Vladimir and Bock, Elisabeth},
  title   = {Regulators of neurite outgrowth: role of cell adhesion molecules},
  journal = {Annals of the New York Academy of Sciences},
  volume  = {1014},
  number  = {1},
  pages   = {140--154},
  year    = {2004},
  doi     = {10.1196/annals.1294.015}
}

@article{Spedden2012Elasticity,
  author    = {Spedden, Elise and White, James D. and Naumova, Elena N. and Kaplan, David L. and Staii, Cristian},
  title     = {Elasticity Maps of Living Neurons Measured by Combined Fluorescence and Atomic Force Microscopy},
  journal   = {Biophysical Journal},
  volume    = {103},
  number    = {5},
  pages     = {868--877},
  year      = {2012},
  doi       = {10.1016/j.bpj.2012.08.005}
}

@article{Spedden2014Asymmetry,
  author    = {Spedden, Elise and Wiens, Matthew R. and Demirel, Melik C. and Staii, Cristian},
  title     = {Effects of Surface Asymmetry on Neuronal Growth},
  journal   = {PLoS ONE},
  volume    = {9},
  number    = {9},
  pages     = {e106709},
  year      = {2014},
  publisher = {Public Library of Science},
  doi       = {10.1371/journal.pone.0106709},
  pmid      = {25184796},
  pmcid     = {PMC4153665}
}

@article{pearson2011modeling,
  author    = {Pearson, Y. E. and Castronovo, E. and Lindsley, T. A. and Drew, D. A.},
  title     = {Mathematical modeling of axonal formation. Part I: Geometry},
  journal   = {Bulletin of Mathematical Biology},
  year      = {2011},
  volume    = {73},
  number    = {12},
  pages     = {2837--2864},
  doi       = {10.1007/s11538-011-9648-2},
  pmid      = {21390561}
}

@article{gladkov2017design,
  author    = {Gladkov, A. and Pigareva, Y. and Kutyina, D. and Kolpakov, V. and Bukatin, A. and Mukhina, I. and Kazantsev, V. and Pimashkin, A.},
  title     = {Design of Cultured Neuron Networks in vitro with Predefined Connectivity Using Asymmetric Microfluidic Channels},
  journal   = {Scientific Reports},
  year      = {2017},
  volume    = {7},
  number    = {1},
  pages     = {15625},
  doi       = {10.1038/s41598-017-15506-2},
  pmid      = {29142321},
  pmcid     = {PMC5688062}
}

@article{kundu2013superimposed,
  author    = {Kundu, A. and Micholt, L. and Friedrich, S. and Rand, D. R. and Bartic, C. and Braeken, D. and Levchenko, A.},
  title     = {Superimposed topographic and chemical cues synergistically guide neurite outgrowth},
  journal   = {Lab on a Chip},
  year      = {2013},
  volume    = {13},
  number    = {15},
  pages     = {3070--3081},
  doi       = {10.1039/c3lc50174d},
  pmid      = {23752939},
  pmcid     = {PMC3820293}
}

@article{ishihara2011assay,
  author    = {Ishihara, M. and Mochizuki-Oda, N. and Iwatsuki, K. and Kishima, H. and Iwamoto, Y. and Ohnishi, Y. and Umegaki, M. and Yoshimine, T.},
  title     = {A new three-dimensional axonal outgrowth assay for central nervous system regeneration},
  journal   = {Journal of Neuroscience Methods},
  year      = {2011},
  volume    = {198},
  number    = {2},
  pages     = {181--186},
  doi       = {10.1016/j.jneumeth.2011.03.020},
  pmid      = {21459107}
}

@article{beighley2012alignment,
  author    = {Beighley, R. and Spedden, E. and Sekeroglu, K. and Atherton, T. and Demirel, M. C. and Staii, C.},
  title     = {Neuronal alignment on asymmetric textured surfaces},
  journal   = {Applied Physics Letters},
  year      = {2012},
  volume    = {101},
  number    = {14},
  pages     = {143701},
  doi       = {10.1063/1.4755837},
  pmid      = {23112350},
  pmcid     = {PMC3477179}
}

@book{strogatz2015nonlinear,
  author    = {Strogatz, Steven H.},
  title     = {Nonlinear Dynamics and Chaos: With Applications to Physics, Biology, Chemistry, and Engineering},
  year      = {2015},
  publisher = {Westview Press},
  address   = {Boulder, CO}
}

@article{Lin2020,
    author = {Lin, Ji  and Li, Xiaokeng  and Yin, Jun  and Qian, Jin },
    title = {Effect of Cyclic Stretch on Neuron Reorientation and Axon Outgrowth},
    journal = {Frontiers in Bioengineering and Biotechnology},
    volume = {8},
    year = {2020},
    URL={https://www.frontiersin.org/journals/bioengineering-and-biotechnology/articles/10.3389/fbioe.2020.597867},
    doi = {10.3389/fbioe.2020.597867},
    issn = {2296-4185}
}

@article{Mahar2018,
    author = {Mahar, Marcus and Cavalli, Valeria},
    date = {2018/06/01},
    doi = {10.1038/s41583-018-0001-8},
    id = {Mahar2018},
    isbn = {1471-0048},
    journal = {Nature Reviews Neuroscience},
    number = {6},
    pages = {323--337},
    title = {Intrinsic mechanisms of neuronal axon regeneration},
    url = {https://doi.org/10.1038/s41583-018-0001-8},
    volume = {19},
    year = {2018}
}

@article{Montanino2018,
    author = {Montanino, Annaclaudia  and Kleiven, Svein},
    title = {Utilizing a Structural Mechanics Approach to Assess the Primary Effects of Injury Loads Onto the Axon and Its Components},
    journal = {Frontiers in Neurology},
    volume = {9},
    year = {2018},
    url ={https://www.frontiersin.org/journals/neurology/articles/10.3389/fneur.2018.00643},
    doi = {10.3389/fneur.2018.00643},
    issn = {1664-2295}
}

@book{Murray1993,
    author = {Murray, James D.},
    title = {Mathematical Biology},
    publisher = {Springer-Verlag, Berlin},
    year = {2013},
    doi = {https://doi.org/10.1007/978-3-662-08542-4},
    edition = {2nd},
}

@article{Oliveri2021,
  title = {Theory for Durotactic Axon Guidance},
  author = {Oliveri, Hadrien and Franze, Kristian and Goriely, Alain},
  journal = {Phys. Rev. Lett.},
  volume = {126},
  issue = {11},
  pages = {118101},
  numpages = {6},
  year = {2021},
  month = {Mar},
  publisher = {American Physical Society},
  doi = {10.1103/PhysRevLett.126.118101},
  url = {https://link.aps.org/doi/10.1103/PhysRevLett.126.118101}
}

@book{Striedter2016,
    author = {Striedter, GF},
    title = {Neurobiology: a functional approach},
    publisher = {Oxford University Press, Oxford},
    year = {2016},
}

@article{Simpson2009,
  title   = {Theoretical Models of Neural Circuit Development},
  author  = {Simpson, H. D. and Mortimer, D. and Goodhill, G. J.},
  journal = {Curr. Top. Dev. Biol.},
  volume  = {87},
  pages   = {1--51},
  year    = {2009},
  doi     = {10.1016/S0070-2153(09)01201-0},
  url     = {https://doi.org/10.1016/S0070-2153(09)01201-0}
}

@article{Segev2000,
  author  = {Segev, R. and Ben-Jacob, E.},
  title   = {Generic modeling of chemotactic-based self-wiring of neural networks},
  journal = {Neural Networks},
  volume  = {13},
  number  = {2},
  pages   = {185--199},
  year    = {2000},
  doi     = {10.1016/S0893-6080(99)00084-2}
}

@article{Krottje2007,
  title   = {A Mathematical Framework for Modeling Axon Guidance},
  author  = {Krottje, J. K. and van Ooyen, A.},
  journal = {Bull. Math. Biol.},
  volume  = {69},
  issue   = {1},
  pages   = {3--31},
  year    = {2007},
  doi     = {10.1007/s11538-006-9142-4},
  url     = {https://doi.org/10.1007/s11538-006-9142-4}
}

@article{Odde1996,
  title   = {Stochastic Dynamics of the Nerve Growth Cone and Its Microtubules during Neurite Outgrowth},
  author  = {Odde, D. J. and Tanaka, E. M. and Hawkins, S. S. and Buettner, H. M.},
  journal = {Biotechnol. Bioeng.},
  volume  = {50},
  issue   = {4},
  pages   = {452--461},
  year    = {1996},
  doi     = {10.1002/(SICI)1097-0290(19960520)50:4<452::AID-BIT13>3.0.CO;2-L},
  url     = {https://doi.org/10.1002/(SICI)1097-0290(19960520)50:4<452::AID-BIT13>3.0.CO;2-L}
}

@article{Buettner1994,
  title   = {A Model of Neurite Extension across Regions of Nonpermissive Substrate: Simulations Based on Experimental Measurement of Growth Cone Motility and Filopodial Dynamics},
  author  = {Buettner, H. M. and Pittman, R. N. and Ivins, J. K.},
  journal = {Dev. Biol.},
  volume  = {163},
  issue   = {2},
  pages   = {407--422},
  year    = {1994},
  doi     = {10.1006/dbio.1994.1158},
  url     = {https://doi.org/10.1006/dbio.1994.1158}
}

@article{Buettner1995,
  title   = {Computer Simulation of Nerve Growth Cone Filopodial Dynamics for Visualization and Analysis},
  author  = {Buettner, H. M.},
  journal = {Cell Motil. Cytoskeleton},
  volume  = {32},
  issue   = {3},
  pages   = {187--204},
  year    = {1995},
  doi     = {10.1002/cm.970320304},
  url     = {https://doi.org/10.1002/cm.970320304}
}

@article{Goodhill1998,
  title   = {Axon Guidance: Stretching Gradients to the Limit},
  author  = {Goodhill, G. J. and Baier, H.},
  journal = {Neural Comput.},
  volume  = {10},
  issue   = {3},
  pages   = {521--527},
  year    = {1998},
  doi     = {10.1162/089976698300017638},
  url     = {https://doi.org/10.1162/089976698300017638}
}

@article{Katz1985,
  author  = {Katz, M. J. and Lasek, R. J.},
  title   = {Early axon patterns of the spinal cord: Experiments with a computer},
  journal = {Developmental Biology},
  volume  = {109},
  number  = {1},
  pages   = {140--149},
  year    = {1985},
  doi     = {10.1016/0012-1606(85)90354-9}
}

@article{Mogilner2005,
  title   = {The Physics of Filopodial Protrusion},
  author  = {Mogilner, A. and Rubinstein, B.},
  journal = {Biophys. J.},
  volume  = {89},
  issue   = {2},
  pages   = {782--795},
  year    = {2005},
  doi     = {10.1529/biophysj.104.056515},
  url     = {https://doi.org/10.1529/biophysj.104.056515}
}

@article{Padmanabhan2018,
  title   = {Axon Growth Regulation by a Bistable Molecular Switch},
  author  = {Padmanabhan, P. and Goodhill, G. J.},
  journal = {Proc. R. Soc. B},
  volume  = {285},
  issue   = {1877},
  year    = {2018},
  doi     = {10.1098/rspb.2017.2618},
  url     = {https://doi.org/10.1098/rspb.2017.2618}
}

@article{Hentschel1999,
  title   = {Models of Axon Guidance and Bundling during Development},
  author  = {Hentschel, H. G. and van Ooyen, A.},
  journal = {Proc. Biol. Sci.},
  volume  = {266},
  issue   = {1434},
  pages   = {2231--2238},
  year    = {1999},
  doi     = {10.1098/rspb.1999.0913},
  url     = {https://doi.org/10.1098/rspb.1999.0913}
}

@article{Goodhill2004,
  title   = {Predicting Axonal Response to Molecular Gradients with a Computational Model of Filopodial Dynamics},
  author  = {Goodhill, G. J. and Gu, M. and Urbach, J. S.},
  journal = {Neural Comput.},
  volume  = {16},
  issue   = {11},
  pages   = {2221--2243},
  year    = {2004},
  doi     = {10.1162/0899766041941934},
  url     = {https://doi.org/10.1162/0899766041941934}
}

@article{Maskery2005,
  title   = {Deterministic and Stochastic Elements of Axonal Guidance},
  author  = {Maskery, S. and Shinbrot, T.},
  journal = {Annu. Rev. Biomed. Eng.},
  volume  = {7},
  pages   = {187--221},
  year    = {2005},
  doi     = {10.1146/annurev.bioeng.7.060804.100446},
  url     = {https://doi.org/10.1146/annurev.bioeng.7.060804.100446}
}

@article{Betz2006,
  title   = {Neuronal Growth: A Bistable Stochastic Process},
  author  = {Betz, T. and Lim, D. and Kas, J. A.},
  journal = {Phys. Rev. Lett.},
  volume  = {96},
  issue   = {9},
  pages   = {098103},
  year    = {2006},
  doi     = {10.1103/PhysRevLett.96.098103},
  url     = {https://doi.org/10.1103/PhysRevLett.96.098103}
}

@article{Cheng2025,
  title = {Feedback-Driven Dynamical Model for Axonal Extension on Parallel Micropatterns},
  author = {Cheng, K. and Kumarasinghe, U. and Staii, C.},
  journal = {Biomimetics},
  year = {2025},
  volume = {10},
  article-number = {456},
  doi = {10.3390/biomimetics10070456},
  url = {https://doi.org/10.3390/biomimetics10070456}
}

@article{Notbohm2016,
  title = {Cellular contraction and polarization drive collective cellular motion},
  author = {Notbohm, J. and Banerjee, S. and Utuje, K. J. C. and Gweon, B. and Jang, H. and Park, Y. and Shin, J. and Butler, J. P. and Fredberg, J. J. and Marchetti, M. C.},
  journal = {Biophys. J.},
  year = {2016},
  volume = {110},
  issue = {12},
  pages = {2729--2738},
  doi = {10.1016/j.bpj.2016.05.019},
  url = {https://doi.org/10.1016/j.bpj.2016.05.019}
}

@article{Banerjee2011,
  title = {Substrate rigidity deforms and polarizes active gels},
  author = {Banerjee, S. and Marchetti, M. C.},
  journal = {EPL (Europhys. Lett.)},
  year = {2011},
  volume = {96},
  article-number = {28003},
  doi = {10.1209/0295-5075/96/28003},
  url = {https://doi.org/10.1209/0295-5075/96/28003}
}

@article{Dokukina2010,
  title = {A model of fibroblast motility on substrates with different rigidities},
  author = {Dokukina, I. V. and Gracheva, M. E.},
  journal = {Biophys. J.},
  year = {2010},
  volume = {98},
  issue = {12},
  pages = {2794--2803},
  doi = {10.1016/j.bpj.2010.03.026},
  url = {https://doi.org/10.1016/j.bpj.2010.03.026}
}

@article{Rubinstein2009,
  title = {Actin-myosin viscoelastic flow in the keratocyte lamellipod},
  author = {Rubinstein, B. and Fournier, M. F. and Jacobson, K. and Verkhovsky, A. B. and Mogilner, A.},
  journal = {Biophys. J.},
  year = {2009},
  volume = {97},
  issue = {7},
  pages = {1853--1863},
  doi = {10.1016/j.bpj.2009.07.020},
  url = {https://doi.org/10.1016/j.bpj.2009.07.020}
}

@article{Barnhart2011,
  title = {An adhesion-dependent switch between mechanisms that determine motile cell shape},
  author = {Barnhart, E. L. and Lee, K. C. and Keren, K. and Mogilner, A. and Theriot, J. A.},
  journal = {PLoS Biol.},
  year = {2011},
  volume = {9},
  issue = {5},
  pages = {e1001059},
  doi = {10.1371/journal.pbio.1001059},
  url = {https://doi.org/10.1371/journal.pbio.1001059}
}

\end{document}